\RequirePackage{fix-cm}
\documentclass[smallextended]{svjour3}       % onecolumn (second format)
\smartqed  % flush right qed marks, e.g. at end of proof
\usepackage{graphicx}
\usepackage{mathptmx}      % use Times fonts if available on your TeX system
\usepackage{natbib}
\begin{document}

%% ------------------------------------------------------------------------ %%
\title{Study of high energy phenomena from near space using low-cost meteorological balloons
}


\titlerunning{Space exploration with meteorological balloons}        % if too long for running head

\author{Sandip K. Chakrabarti \and
        Ritabrata Sarkar \and
        Debashis Bhowmick \and
        Arnab Bhattacharya
}

\authorrunning{Chakrabarti et~al.} % if too long for running head

\institute{Sandip K. Chakrabarti 
           \and Ritabrata Sarkar 
           \and Debashis Bhowmick 
           \and Arnab Bhattacharya 
           \at Indian Centre for Space Physics, 43 Chalantika, 
           Garia Station Rd., Kolkata 700084\\
           Tel.: +91-33-24366003\\
           Fax: +91-33-24622153 Ext. 28\\
           \email{ritabrata.s@gmail.com}
           \and Sandip K. Chakrabarti 
           \at S.N. Bose National Centre for Basic Sciences,
           JD Block, Salt Lake, Kolkata 700097 \\
           Tel.: +91-33-23355706\\
           Fax: +91-33-23353477\\
           \email{sandipchakrabarti9@gmail.com}
}

\date{}

\maketitle

%% ------------------------------------------------------------------------ %%
\begin{abstract}
Indian Centre for Space Physics has taken a novel strategy to study 
low energy cosmic rays and astrophysical X-ray sources which involve very light weight 
payloads up to about five kilograms on board a single or multiple balloons which are usually used for 
meteorological purposes. The mission duration could be anywhere from 3-12 hours.
Our strategy provides extreme flexibility in mission preparation and
its operation using a very economical budget. There are several limitations but our innovative
approach has been able to extract significant amount of scientific data out of these missions. 
So far, over one hundred missions have been completed by us to near space and a wealth of data has been collected. 
The payloads are recovered and are used again. Scientific data is stored on board computer
and the atmospheric data or payload location is sent to ground in real time. 
Since each mission is different, we present here the general strategy for a typical payload
and provide some results we obtained in some of these missions.


\end{abstract}

%% ------------------------------------------------------------------------ %%
\section{Introduction}
\label{intro}
Study of the Universe in X-rays is a very important regime of astronomical 
observations. Since the atmosphere surrounding the earth obscures X-ray, 
we need to reach above it. For more than hundred years, balloons 
are being used for this purpose to study Cosmic Ray (CR) and high energy 
phenomena of the Universe. More expensive satellite missions 
with on board payloads provide astronomers a platform to carry out
experiments which can do longer observations. These missions are expensive and
require a very long preparation time.

With the advent of miniaturization of instruments, and at the same time 
improvement of technology to produce low cost balloons, it has become now possible to
investigate the high energy universe using very light-weight payloads.
Successful recovery of payloads make the instruments reusable rendering the cost even lower and within
the reach of colleges and university budgets. While our integrated payloads are
less than about $5$ kg, some of the science-return has been rewarding. Because of light-weight,
pointing had to be sacrificed, but our innovative approach allows us to separate photons
from each object of interest. Certainly, some scientific goals which require longer
observations, larger areas or precession pointing are sacrificed, but repeatability at a low cost 
enables us to study very long time scale variability (such as cosmic ray variation with solar activity
for several years) which are not possible by a single satellite mission \citep{sark17}. In addition, our method
allows us to evaluate, test and improve each payload in a near space environment before being used in space. 
Several preliminary reports of our activities are given in \citet{chak11, chak13, chak14, chak15}.

In this paper, we describe the overall mission strategy, configuration of a
typical payload and give examples of typical results obtained by such a payload.
In the next Section (\S \ref{sec:miss}), we present the overview of such 
missions. The detailed description of the payload components and their
properties are discussed in \S \ref{sec:payl}. Then in \S \ref{sec:calib} we 
discuss the test and calibration of the payloads. In \S \ref{sec:res}
we show and discuss typical results, namely, spectrum, light curve and other 
byproducts from such missions. Finally, in \S 6, we make concluding remarks.

%% ------------------------------------------------------------------------ %%
\section{Mission description}
\label{sec:miss}

\subsection{General philosophy}
\label{ssec:phil}
Our motivation is to carry payloads which are no more than about 5 kilograms and to use
light weight (1 to 2 kg category) rubber or plastic balloons of about 4000 cubic meters
category which are generally used for meteorological purposes. Our science goal is
to concentrate on the X-ray sky, though similar fruitful science could be done in
any wave band. Our requirements are therefore very stringent in the sense that further in the
sky we send our payloads, lesser is the absorption due to the atmosphere and lower is the 
detectable energy band in X-rays. To this effect, we typically fly from 35 to 42 km.

Each payload, apart from the X-ray instrument in question, must have the data acquisition
system, an attitude measure measurement system, GPS system transmitting the instantaneous
locations to the ground, parachutes (for rubber balloons), one or more optional video cameras, 
an optional sun-sensor etc. If the payload is heavier than 2.5 kg, we use two balloons with suitable 
(to be discussed below) lifts. To have higher buoyancy, we use hydrogen gas for filling the 
balloons. So far, we flew our Dignity series of payloads which are used for Cosmic rays 
and X-ray Universe, 101 times and except for a couple of missions in early days when 
our expertise were limited, all the payloads have been successfully recovered.

We designed the balloon borne payload to be compact, light-weight, reusable
and most importantly of low-cost. Indeed, typically we are able to fly 10-15 times per year in 
typically two preferred seasons when easier recovery is possible. 
Construction of a high-altitude balloon-borne payload must take care of 
the following design constraints. At higher altitudes, the temperature 
becomes cryogenic. Under that environment, most of the general 
purpose electronic devices and power supplies are expected to fail
and all circuits calibrations on ground are changed. The response of the
detectors would also vary. Thus the payload box must be provided with a temperature
shielding and the system should be tested to withhold the pressure variation as the
mission reaches from $\sim 1000$ mbar to $\sim 2$ mbar. Since the mission is 
expected to be of few hours, and could also be at night time, solar cells are   
not used.

Styrofoam (thermocol) is a very common widely available heat
insulator with a thermal conductivity of only $0.029$ W/m$^{\circ}$C and is of
considerably low-cost. Thermocol is a smooth and 
soft, light weight, material enabling us to curve out areas where specific payloads or circuit boards
are to be placed. It has a good amount of strength, and yet, has 
the ability to absorb some mechanical shock of possible impact with earth during landing.
Because of these features, it is the perfect choice for our payloads. All
the electronics and detector are also covered with separate thermocol boxes
for extra protection. Typically, our payloads remain at $10-15^{\circ}$C 
inside the box throughout the journey. 

During our long experience with payload boxes of various shapes and sizes, 
judging from the data on turbulence experienced in its journey, especially during crossing 
the tropopause, we found it better to choose the shape of the payload to be cylindrical.
The base of the cylinder is flat where electronics components and power supplies rest.
On the top surface we place the collimator for the X-ray detector, whenever needed.
It may also contain GPS antennas, sun sensor and video camera. Side walls may also
have cameras whenever needed. Bottom surface has an extra protection of semi-hard 
plastic cylinders to absorb shocks during landing.

\subsection{Dependence of Balloons and flight-times on Mission goals}
\label{ssec:carr}
One of the major flexibilities of our method is that the launching location 
need not be fixed. Indeed, we shift the launching location among one of the $4-5$
chosen places, depending on the wind pattern and the desired landing location.
The landing location strongly depends on the wind pattern and changes 
significantly from day to day, and most certainly, also on whether both balloons burst 
or a single balloon bursts. Our mission purpose dictates the time of the lift off
and that also shifts the landing location within a given day. Mission aim
decides the payload weight. Typically, for lighter weight (combined weight of 
the payload components with parachute less than about $2.5$ kg) we fly the
missions with a single meteorological balloon. The flight duration is short 
(up to about $3$ hours) as the balloon starts descending right after the burst. 
However when the payload is of intermediate weight (e.g., $3-3.5$ kg), we send 
two balloons tied up with lifts in such a way that 
one bursts first and the other brings the payload down slowly, and even float at a 
near neutrality height for a considerable time. These missions are of longest
durations. We experimented with a case of $12$ hours of flight with two balloons. 
However, typical duration is about $5-6$ hours. The landing location has to be predicted
very carefully depending on the lifts given and whether or not near neutrality is
achieved. Of course, when GPS is attached, it becomes easier for the recovery team to
collect the payload in real time as it lands. 

When the payload is heavier, say $4$ kg or more, we either send two rubber balloons with 
appropriate lifts so that both burst simultaneously and the trajectory is close to that
predicted for a single balloon. Even when one bursts, the heavy payload is brought
down by the other, merely an hour or so late. So the flight duration is about $3-4.5$ hours
in this case. However, if we are interested to have longer duration data at 
greater height (say above 30km), we prefer to use a single polythene balloon which can
stay about $4-5$ hours in flight with about an hour above $30$ km. We use TIFR, Hyderabad
made $4000$ cubic meter category balloons for this purpose. Since a polythene balloon 
does not burst as a rubber balloon, and also does not require any parachute, 
its trajectory is quite different from those flights having rubber balloons. These balloons also 
require a larger launching ground. More than two smaller 
polythene balloons have been tried by us also. However, the predictability of 
such balloons is not high and recovery becomes more complex. So, we do not recommend 
using more than two balloons without ejection mechanism, unless there are 
specific requirements for such a configuration. They tend to go to lower height also, mainly due to 
uneven load on each balloon during air turbulence.

\begin{figure}[h]
  \centering
  \includegraphics[width=0.32\textwidth]{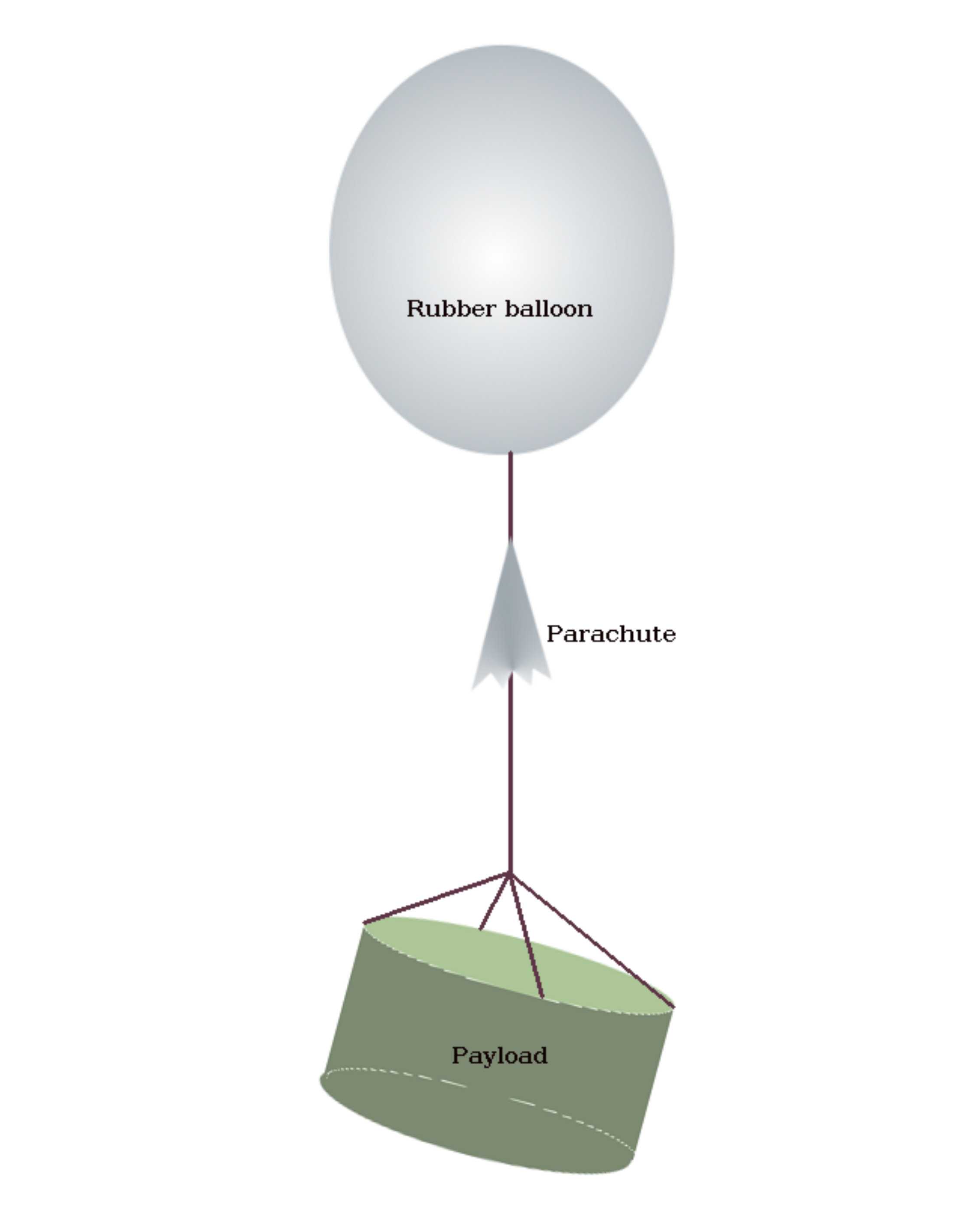}
  \includegraphics[width=0.32\textwidth]{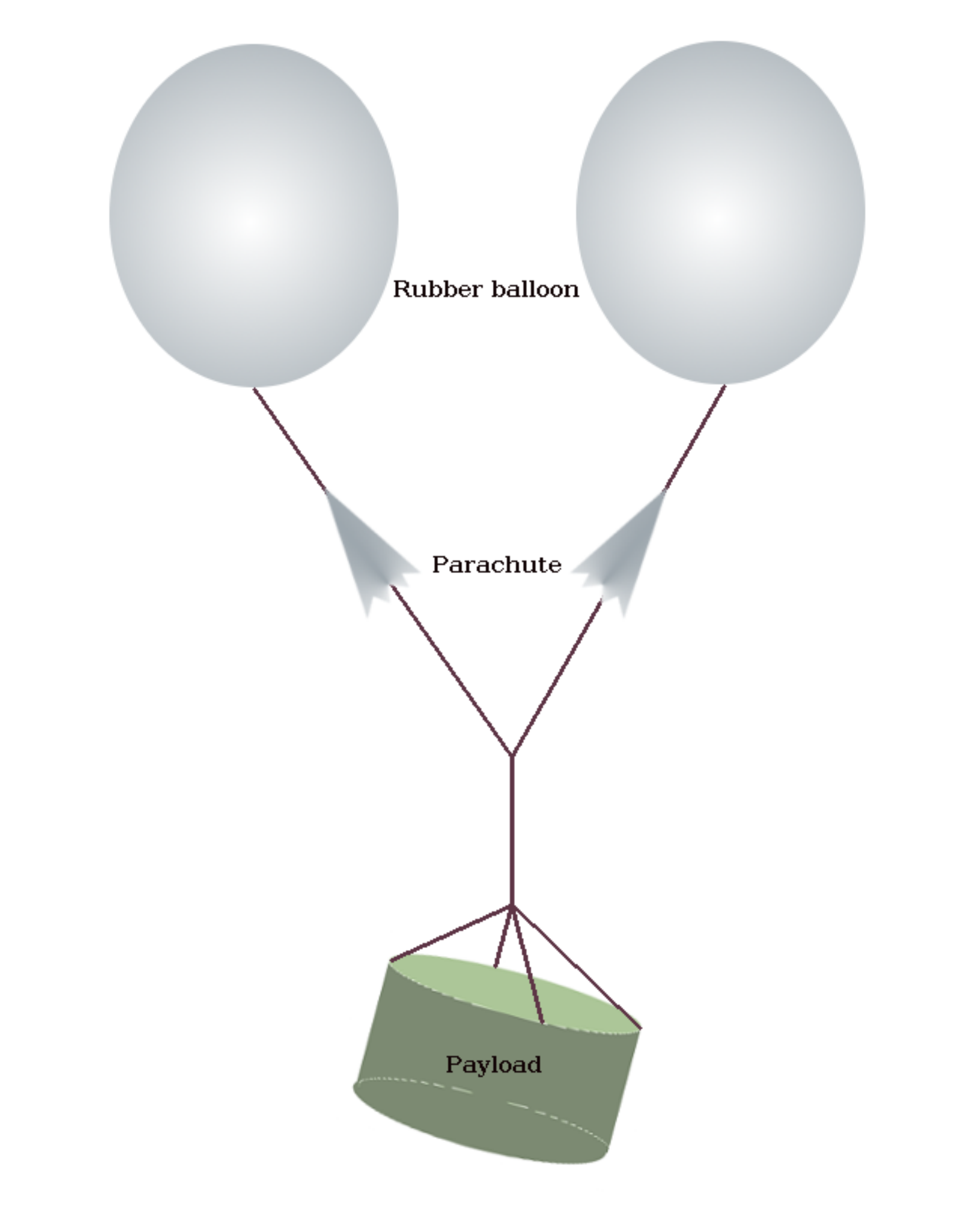}
  \includegraphics[width=0.32\textwidth]{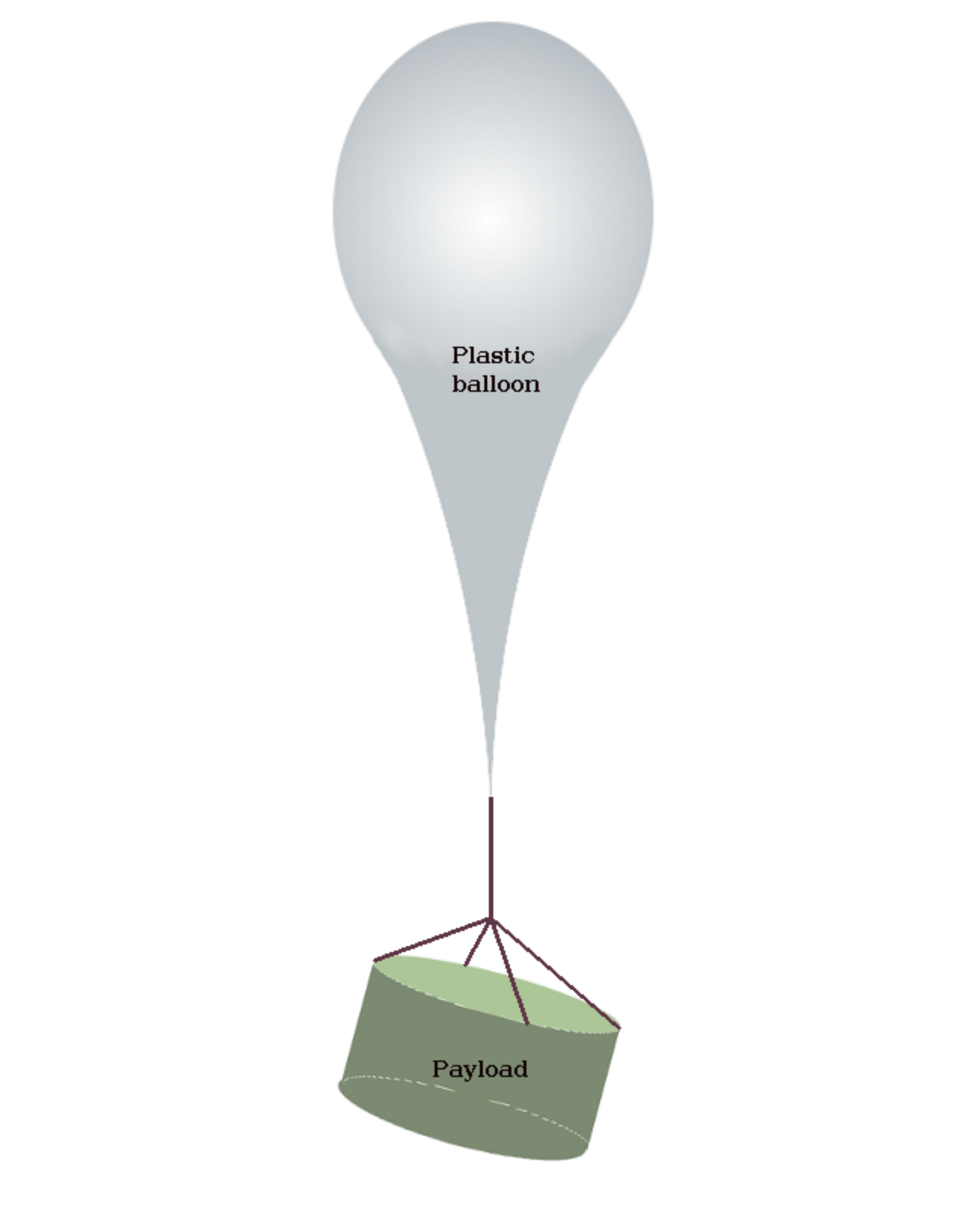}
  \caption{Different carrier configurations used in the flights. Left: single 
  rubber balloon ($\sim 10$ ft diameter on the ground); 
middle: double rubber balloon; right: About 100 ft long polythene balloon.}
   \label{fig:flttyp}
\end{figure}

The schematic picture of different types of carrier balloons and their
combinations to carry the payload are shown in Fig. \ref{fig:flttyp}.
In case of the stretchable rubber balloons the payload is suspended by a rope 
at the end of the balloon with a parachute inserted in between the balloons
and the payload.  

\begin{figure}[h]
  \centering
  \includegraphics[width=0.32\textwidth]{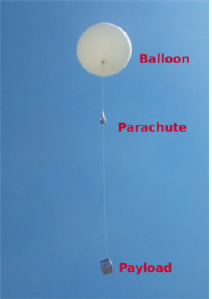}
  \includegraphics[width=0.32\textwidth]{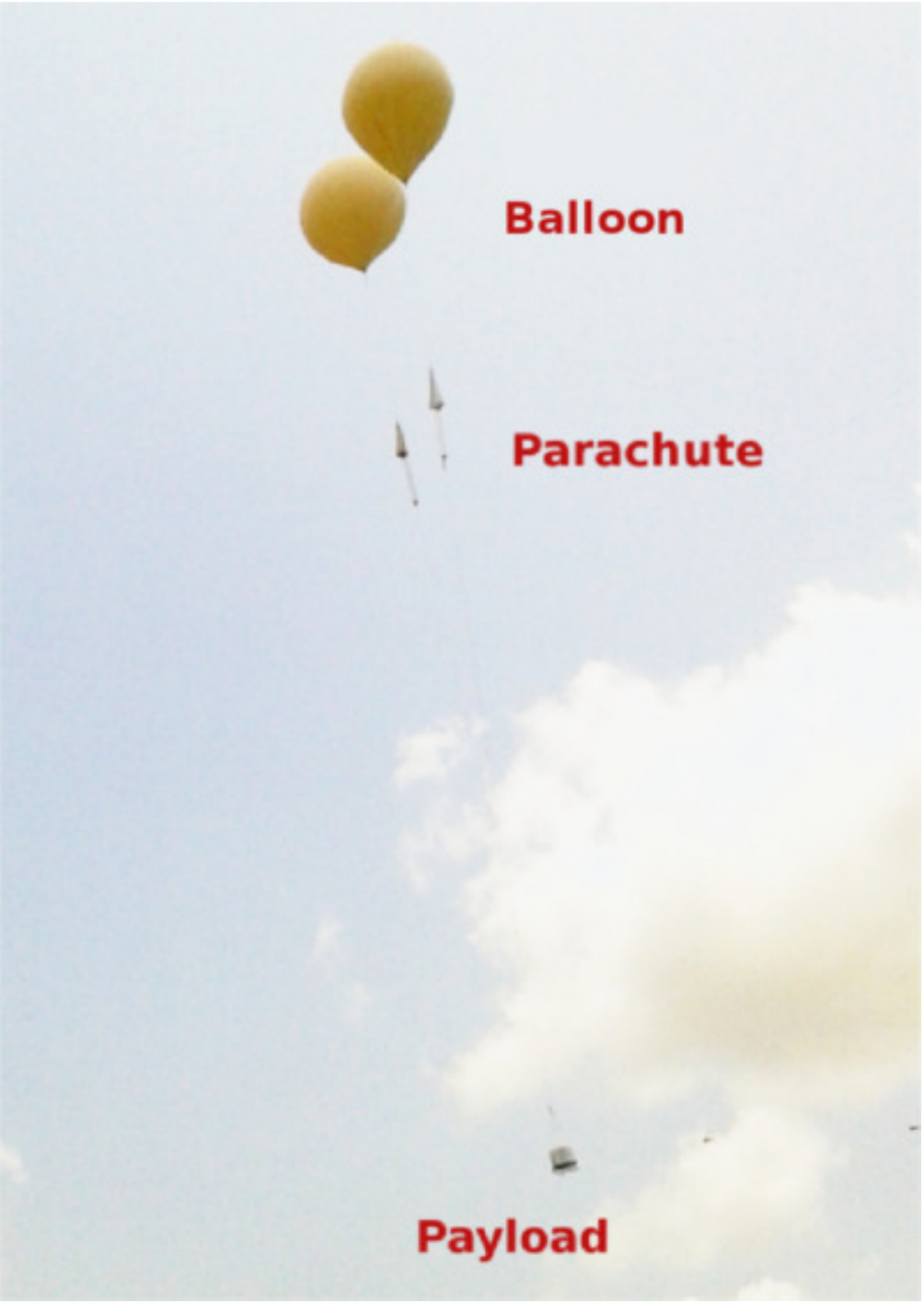}
  \includegraphics[width=0.32\textwidth]{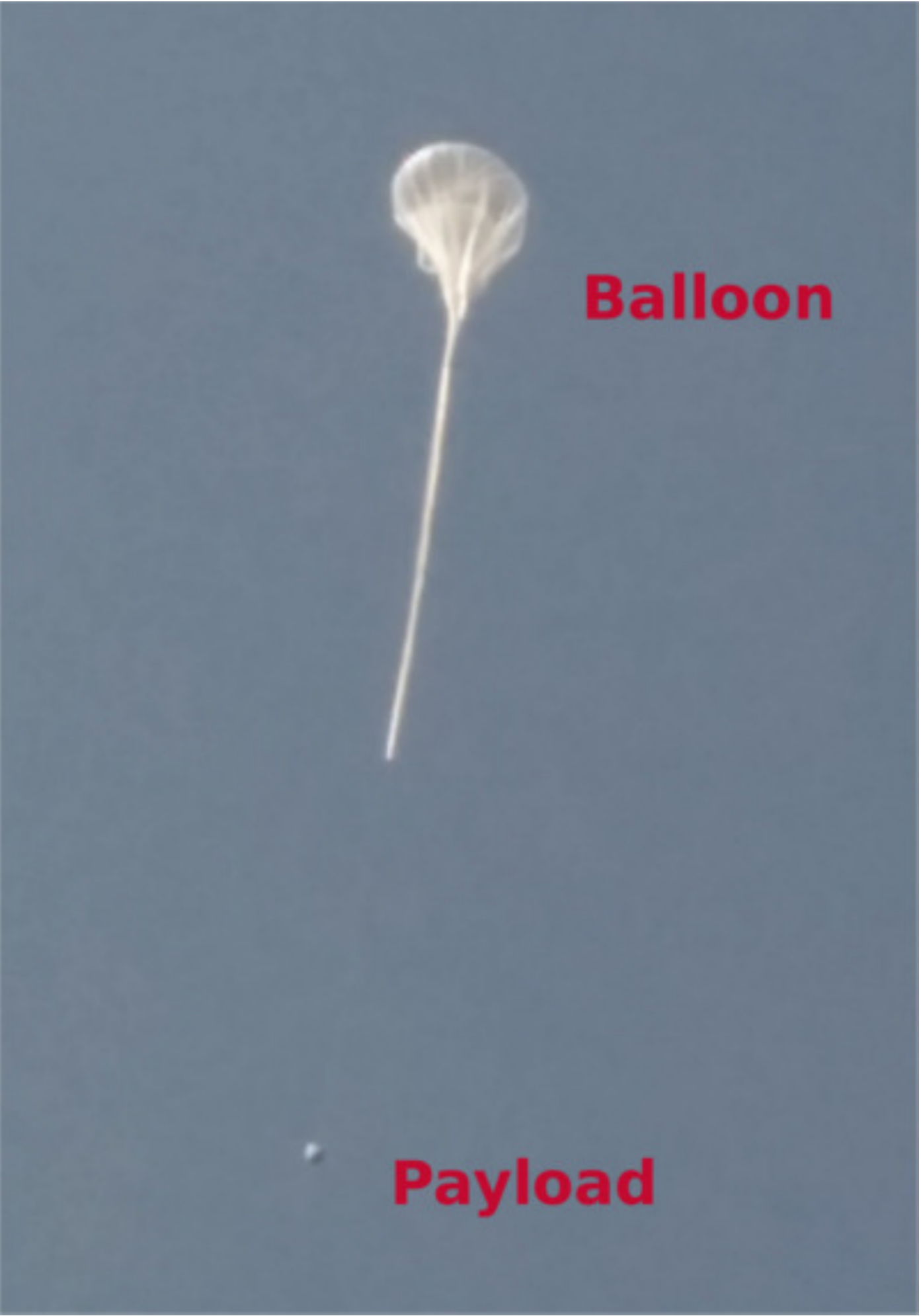}
\caption{Balloons and payload photos just after launch. Left: single 
rubber balloon; middle: double rubber balloon; right: bigger polythene
balloon. Individual parachutes are attached to  each of the double balloons
to avoid possible crash landing.}
   \label{fig:balpic}
\end{figure}

\begin{figure}[h]
  \centering
  \includegraphics[width=0.4\textwidth]{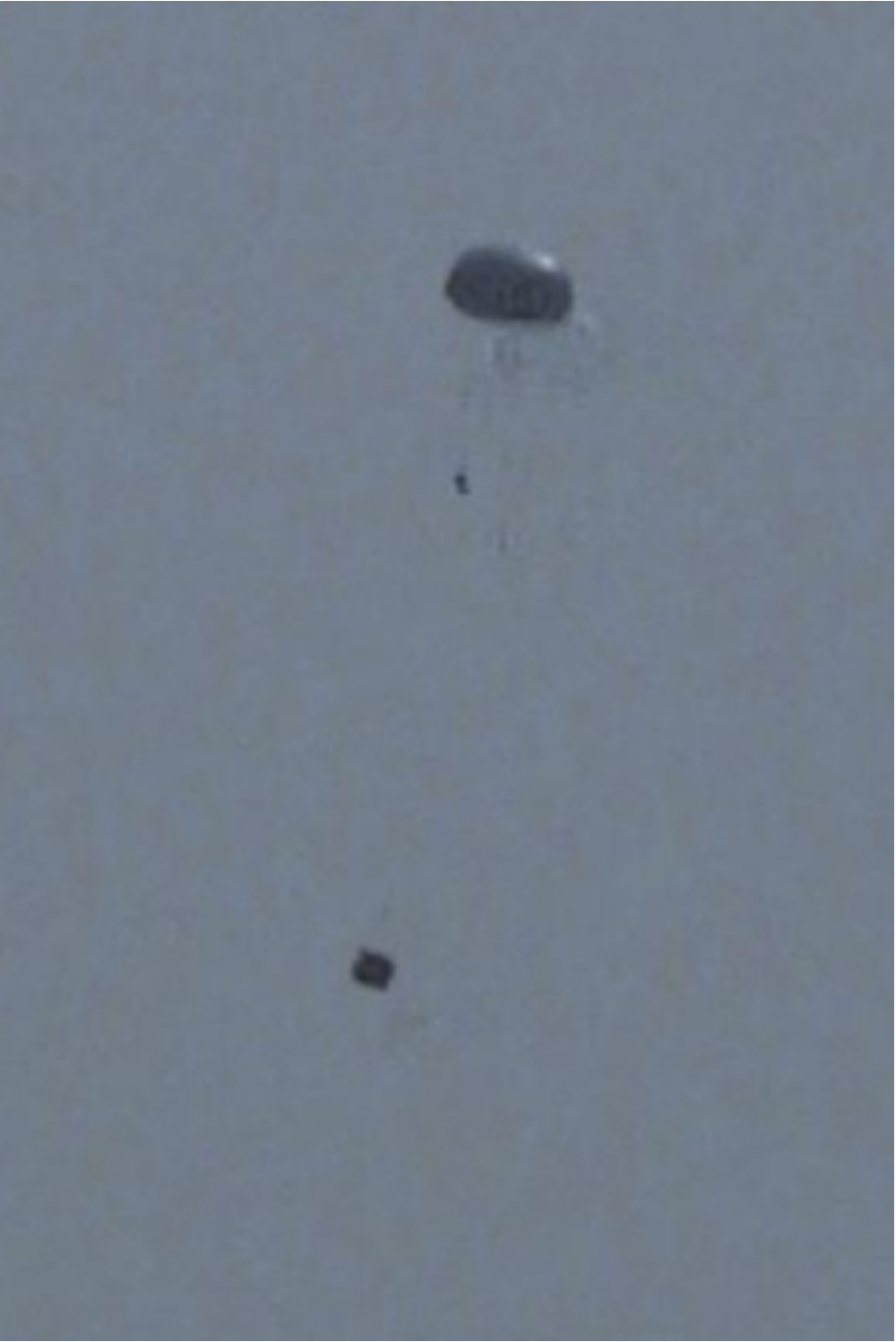}
  \caption{Example of a Parachute (D17) carrying the payload back to the ground.}
   \label{fig:para}
\end{figure}

In Fig. \ref{fig:balpic} we show the real pictures of three kinds of carrier 
combinations.
A picture of typical landing with the parachute is shown in Fig. \ref{fig:para}.

\subsection{Trajectory}
\label{ssec:traj}
The overall flight path depends on the duration of the flight and the wind
condition and we can roughly predict the path using one of the numerous balloon 
flight path predictors. They generally predict the landing location for a single
balloon, but in complex situations such as ours, we use our own software
to predict the location depending on the lifts given at launch.

\begin{figure}[h]
  \centering
  \includegraphics[width=0.6\textwidth]{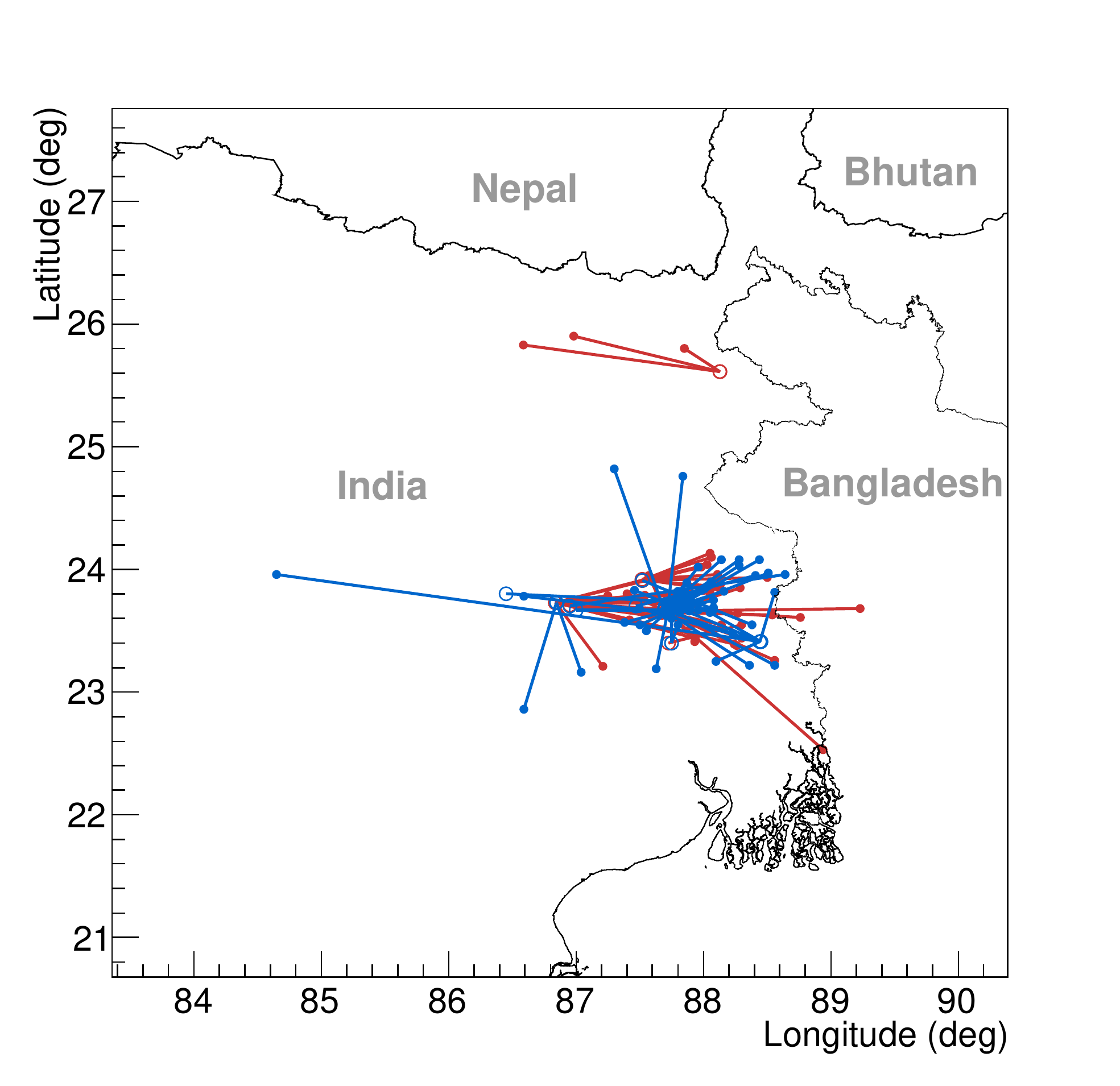}
  \caption{Distribution of the launching (open circles) and landing locations (filled circles) 
of the payloads. Missions during the Jul-Feb are shown by red and those during the
   Mar-Jun are given by blue.}
   \label{fig:locs}
\end{figure}

Single balloon landing prediction routines such as the Cambridge University 
\citep{habhub} tool helps us to choose suitable 
launching location in order that the landing is at a desired place for safer payload 
recovery. Typically, in our geographical location of launching,
namely, Bolpur, West Bengal (India) having (Latitude: 23.67N, Longitude: 87.68E)
and in its vicinity, there are two windows of opportunity for lunching
per year each of which last for about a month. One is in April-May, in the pre-monsoon season, 
and the other is in October-November, in the post-monsoon season. Of course, we flew at other times as well (Fig. 4). 
In Fig. \ref{fig:locs} we show a plot of the launching and landing locations
for the missions we conducted so far. Only the end points and not the
the actual paths are shown.

We rely on the transmitted GPS data every 30 seconds to get the actual locations.
However, whenever the weight restriction constrains us, we use only the GPS locator 
service through local cell phone network to get the landing location. These allow us 
to have almost a hundred percent recovery. Because of international boundaries nearby,
we restrict to these two launching slots. In future, we plan to shift launching towards
central India to have round the year launching without any problem.

\subsection{Behaviour of Lifts in a single and a double-balloon configuration}
\label{ssec:dbal}

\begin{figure}[h]
  \centering
  \includegraphics[width=0.6\textwidth]{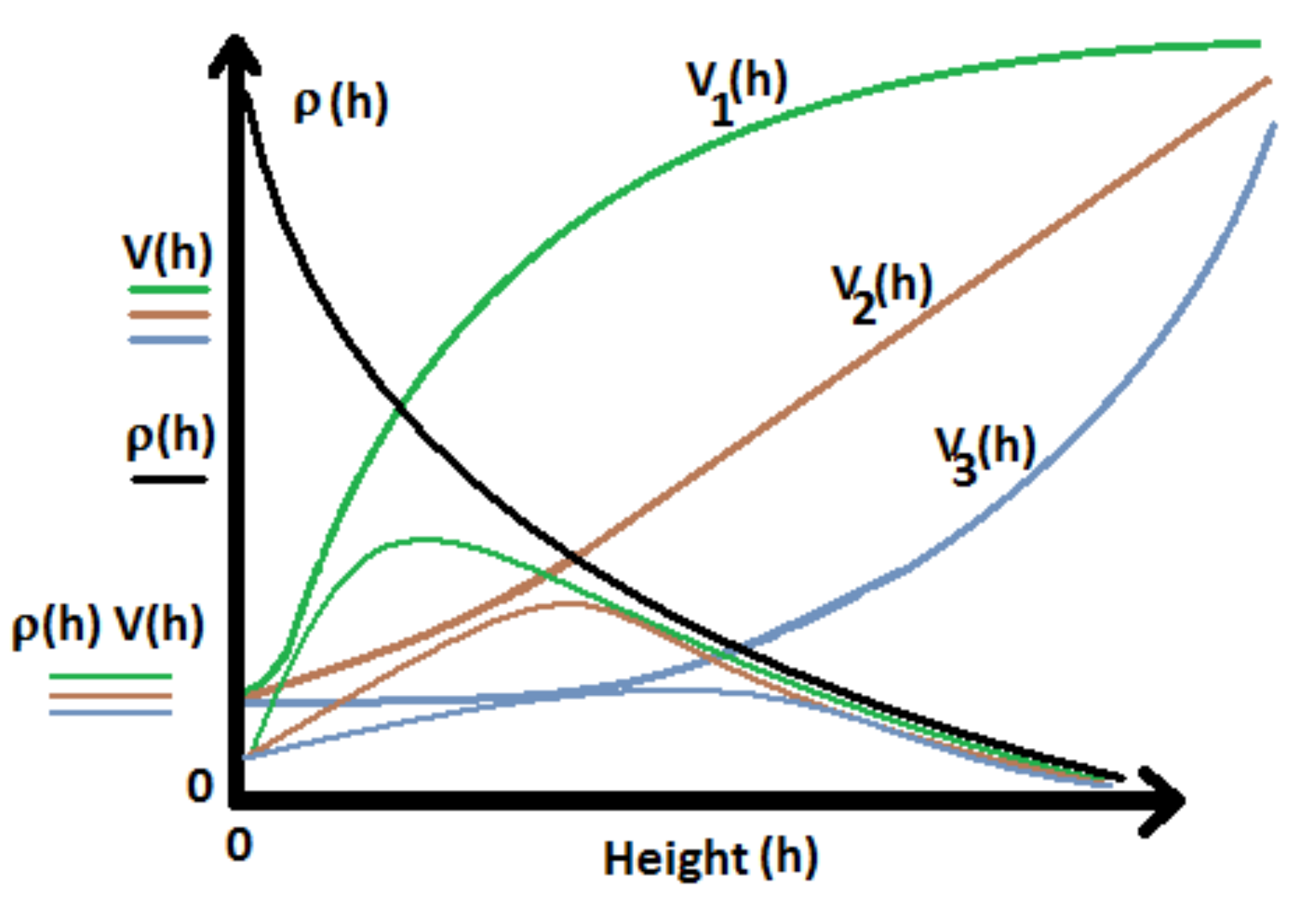}
  \caption{Schematic behaviour of the air density $\rho(h)$, 
balloon volume $V(h)$,  and the lift $\rho(h)V(h)$ as a function of height for three types of expansion of 
the balloon with height. All quantities are in arbitrary units.}
   \label{fig:singrhov}
\end{figure}

\begin{figure}[h]
  \centering
  \includegraphics[width=0.5\textwidth]{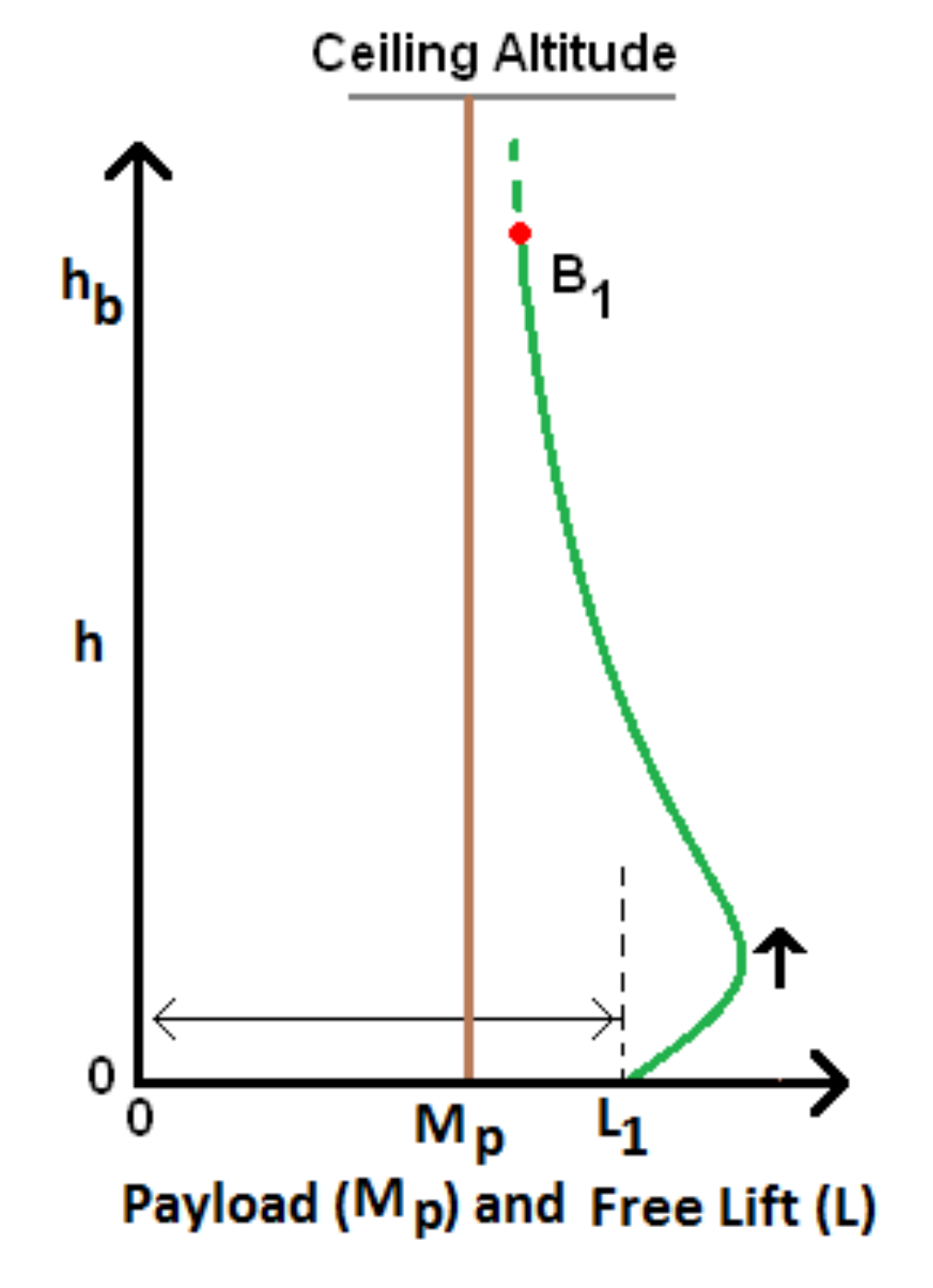}
  \caption{Schematic behaviour of $\rho(h) V(h)$ of a general balloon with altitude showing that the lift
has a maximum at an altitude and then starts decreasing monotonically till the burst height.}
   \label{fig:singbal}
\end{figure}

In a single balloon with the configuration shown in Fig. 1 (left), the effective free
lift is given by
$$
L=\rho(h) V(h)=M_p \frac{a}{g} - M_H,
$$
where, $V(h)$ and $\rho(h)$ are the volume of the balloon 
and the density of air respectively at the instantaneous height $h$;
$M_p$ and $M_H$ are the masses of the payload ($M_p=M_i+m$;
$M_i$= mass of the instrument box; $m$= mass of the parachute) 
and $M_H$= mass of the hydrogen gas inside the balloon. $a$ is the 
acceleration in which the payload lifts off and $g$ is the acceleration due to gravity.

\begin{figure}[h]
  \centering
  \includegraphics[width=0.6\textwidth]{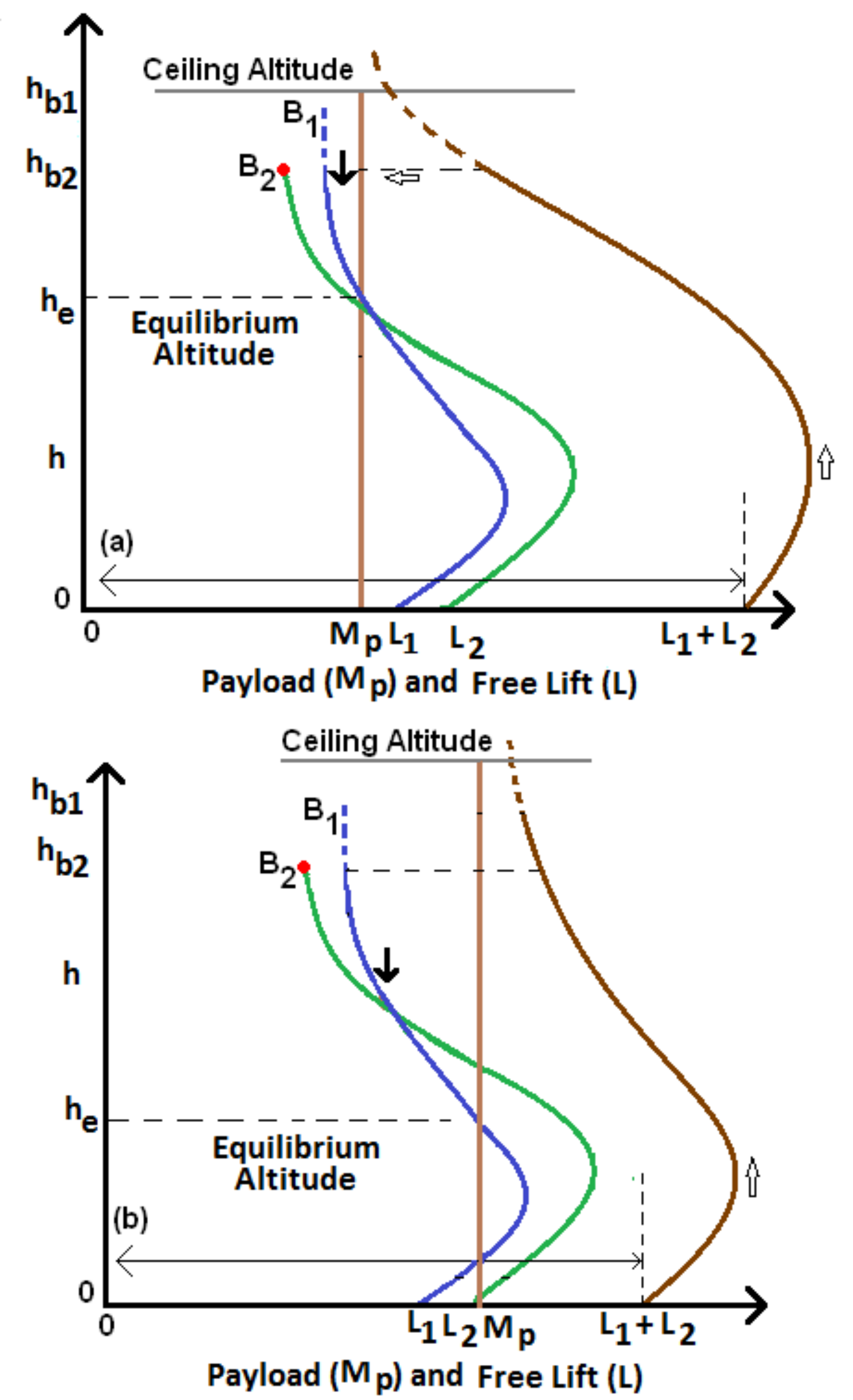}
  \caption{Example of the cases when it is possible to have a neutral buoyancy 
when a single balloon bursts and the other remains intact in a double-balloon
configuration. (top) When both balloons have lifts  much higher than the payload
weight. (bottom) When the lifts are comparable or less than the payload weight,
such as when the payload is massive. See text for details.}
   \label{fig:dbal}
\end{figure}

In Fig. \ref{fig:singrhov}, we show the general trends of
$\rho(h)$ (in black), $V(h)$ for three cases, in green, brown and blue curves
marked by $V_1$, $V_2$ and $V_3$ respectively and $\rho(h) V(h) $ for the 
same three cases (not marked for clarity). In all the cases, there is a height where the lift is 
maximum. Actual height depends on balloon material.  
In Fig. \ref{fig:singbal}, we show this lift once more for a single case
where the height is plotted vertically. In Fig. \ref{fig:dbal} this is shown for
two balloons $B_1$ (blue) and $B_2$ (green)
which are tied together. There could be two situations: In the first case (top),
the net lift $L=L_1+L_2$ is much higher compared to the payload mass $M_p$
(as each of the lifts $L_1$ and $L_2$ is higher than the payload mass $M_p$)
and at a break altitude $h_{b2}$ one balloon, say $B_2$ (green), bursts and $B_1$ has to 
carry the payload alone till its lift is the same as $M_p$ and the balloon reached 
neutrality at an equilibrium altitude $h_e$. With decrease in temperature at nightfall
the lift is reduced and system descends through the maximum lift point to the ground. 
In the second case (bottom), individual lifts are very low, but the highest lift
point still crosses the mass of the payload $M_p$. In this case, the equilibrium height
is very low, and it descends even as the temperature is reduced slightly
(e.g., by the sun set). 

%% ------------------------------------------------------------------------ %%
\section{Payload description}
\label{sec:payl}
The payload of a typical mission includes a Main Measurement Unit (MMU)
which, for example, could be a scintillator type detector or Geiger-M$\rm{\ddot{u}}$ller 
(GM) counter. The MMU must be supported by some other essential/optional 
ancillary equipments like Attitude Measurement Unit (AMU), Video Camera Unit 
(VCU), Sun Sensor (SS) and Global Positioning System (GPS) Unit (GPSU). 
Tracking and recovery of the payload is handled by Positional Information 
Transmission Unit (PITU) and/or simply Position Alert System Unit (PASU) 
using cellphone network (GSM communication) which is automatically
activated upon landing when it is under GSM network coverage. The payload units
and their components along with their purposes are summarized in Table
\ref{tab:payload}. The Power Supply Units (PSU) consist of a stack of rechargeable 
batteries and are generally Li-Po type. Instead of a centrally powered 
system, we put several such units to
increase the chance of receiving some data in case of failure of some PSUs.

\begin{table}[h]
 \begin{center}
   \begin{tabular}{cp{3cm}p{6cm}}
    \hline
    {\bf Unit} & {\bf Component} & {\bf Purpose} \\
    \hline
    \hline
    MMU & Scintillator (2" or 3" NaI, 5" CsI/NaI Phoswich) & Spectral and 
    temporal X-ray and cosmic ray measurement. \\
    & GM counter & Temporal X-ray and cosmic ray measurement. \\
    \hline
    AMU & 9DOF, P-T sensors & Payload attitude measurement in terms of altitude
    and azimuth angle, also includes pressure and temperature measurements. \\
    \hline
    VCU & HD video camera & Optical movie and/or image of the sky, ground/
    carrier balloons/meteorites/sun and planets. \\
    \hline
    SS & Photo sensor & To cross-check if the Sun is inside the detector FoV. \\
    \hline
    GPSU & GPS receiver & Record the flight path using GPS satellites. \\
    \hline
    PASU & GPS tracker & Send location upon landing using GSM network. \\  
    \hline
    PITU & Wireless transmitter & Send on flight positional information
    recursively in predefined time interval. \\
    \hline
   \end{tabular}
   \caption{Payload components and their purpose.}
   \label{tab:payload}
 \end{center}
\end{table}

The SS normal axis and AMU sensor z axis are always parallel to the MMU z axis
(i.e., normal to the detection surface) so that both can see the same object.
The orientation of the video camera or cameras may vary 
depending on the purpose of observation.
The MMU detector is generally attached to the top lid of the payload container
for clear access of all electronics and battery inside the container. 

The overall payload orientation depends on the observation goal we are
interested in and varies with the position of the target object in the sky
during the mission flight and we fix the tilt angle of the payload accordingly
when we suspend the payload from the balloon.

\begin{figure}[h]
  \centering
  \includegraphics[height=0.5\textwidth]{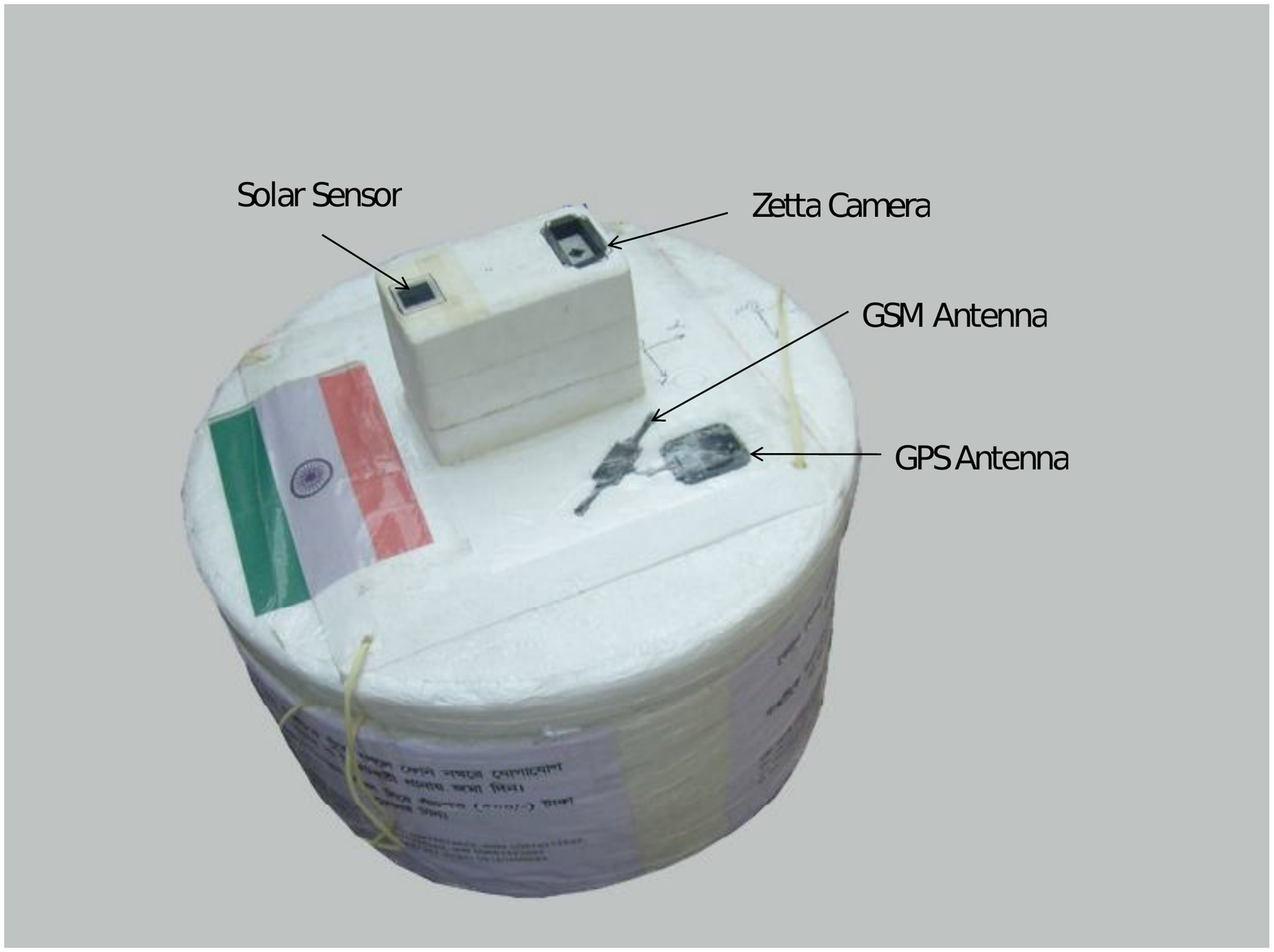}
  \includegraphics[height=0.5\textwidth]{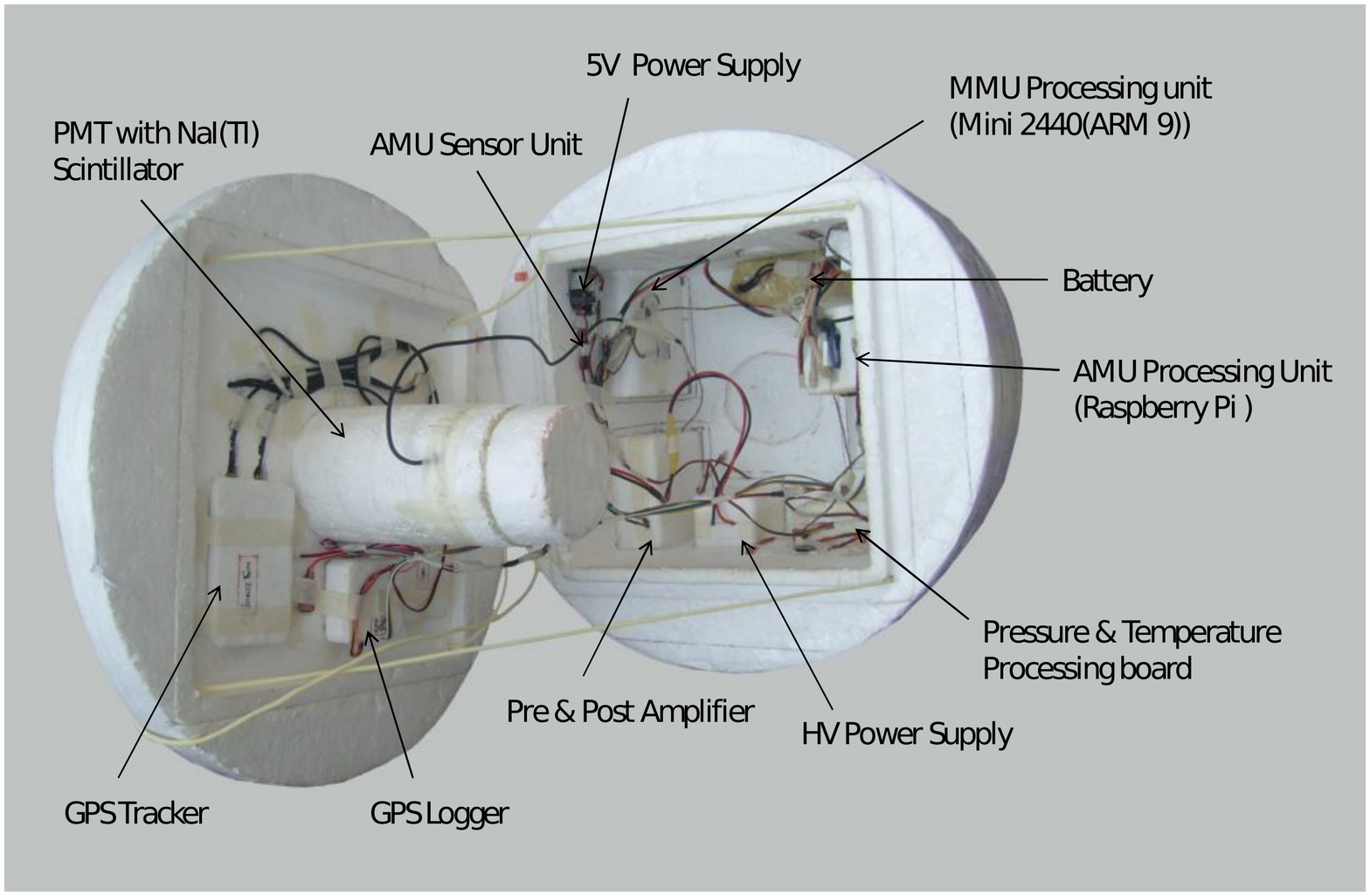}
  \caption{Overall view of a typical payload (top) and the 
internal placements of the main instrument and other components (bottom).}
   \label{fig:payload}
\end{figure}

The GPS and GSM antenna are always on the top lid of the payload container
or at the side depending on the payload tilt angle so that it can see the
sky to be able to receive GPS satellite data. In Fig \ref{fig:payload} 
we present a typical payload as seen from outside (top). We also show the arrangements 
of the main instrument along with the ancillary instruments with the lid open 
(bottom).

\subsection{Main measurement unit}
\label{ssec:mmu}
In ICSP balloon experiments our main goal has been the temporal and spectral study of high energy
radiation, be it from cosmic rays or solar or other extra-terrestrial objects. We therefore keep an 
X-ray detector as the main detector in the payload. This can be changed according to the need. Considering
the goal of our experiment and constraints on energy range, mechanical
robustness required for the mission, weight limitations etc., we chose
scintillator detectors as the candidate. Sometimes to measure the secondary CR
events in the atmosphere we use GM counters as the MMU on board the mission (see, 
Chakrabarti et al. 2011). This measurement unit consists of two main
parts: X-ray detector unit and readout electronics unit to record the events 
from the detector. We discuss these
parts in more detail along with other parts such as collimator, power supply etc.
in the following Section.

\subsubsection{Scintillator detector}
\label{sssec:scin}
The scintillator detector consists of a scintillating crystal optically mounted on 
a Photo-Multiplier Tube (PMT) for the Primary readout of the energy deposition.
Due to efficient energy absorption mechanism, Sodium Iodide (NaI) and Cesium
Iodide (CsI) have been the most useful inorganic scintillator material for hard
X-rays. NaI yields more light and consequently better energy resolution, though
it should be handled more carefully during the packaging since it is extremely
hygroscopic. We use a NaI(Tl) (and in some cases CsI(Na)) crystal hermetically
sealed in an aluminum housing to prevent the crystal quality deterioration.

The PMT is optically coupled directly to the scintillator crystal. The PMT is guarded by
a mu-metal magnetic shield. The scintillator container and mu-metal shield are
sealed together to form a low-mass and light-tight housing for the detector.

The working energy range of the detector depends on the energy deposition
efficiency of the crystal which depends on the crystal material as well as on
its dimension.

We used detectors with different crystal sizes mounted on a suitable PMT according 
to the mission purpose(s). We mostly used cylindrical crystals of NaI(Tl)
and in some cases CsI(Na). Apart from these single crystal scintillator detectors,
sometimes we also used Phoswich detectors where two different kinds of crystals (viz.
NaI(Tl) and CsI(Na)) are optically coupled together with a single PMT. This type of 
scintillator detector is used for better background subtraction and better spectral 
data. A list of detectors with their specifications is given in Table \ref{tab:scnt}.

\begin{table}[h]
 \begin{center}
  \begin{tabular}{lcccc}
   \hline
   Detector & Crystal & Thickness & Diameter & Weight\\
   & material & (cm) & (cm) & (with PMT)(g)\\
   \hline
   Bicron 2M2/2 & NaI & 5.08 & 5.08 & 900\\
   Bicron 3M3/3 & NaI & 7.62 & 7.62 & 1900\\
   Harsaw (3'') & NaI & 7.62 & 7.62 & 2100\\
   Philips (5'') & CsI & 0.6 & 11.60 & 1050\\
   Phoswich (5'') & NaI + CsI & 0.3 (NaI), 2.5 (CsI) & 11.60 & 2700\\
   \hline
  \end{tabular}
  \caption{Properties of the scintillators (NaI(Tl) and CsI(Na)) crystals used
  in different missions.}
  \label{tab:scnt}
 \end{center}
\end{table}

The electronics readout system in the detector module consists of: (i) front-end
electronics with signal amplifiers, (ii) electronics for event pulse
generator (triggering circuit), (iii) peak hold electronics, (iv) digitization
and on board storage system and (v) high voltage DC-DC converter power supply.

\begin{figure}[h]
  \centering
  \includegraphics[width=0.6\textwidth]{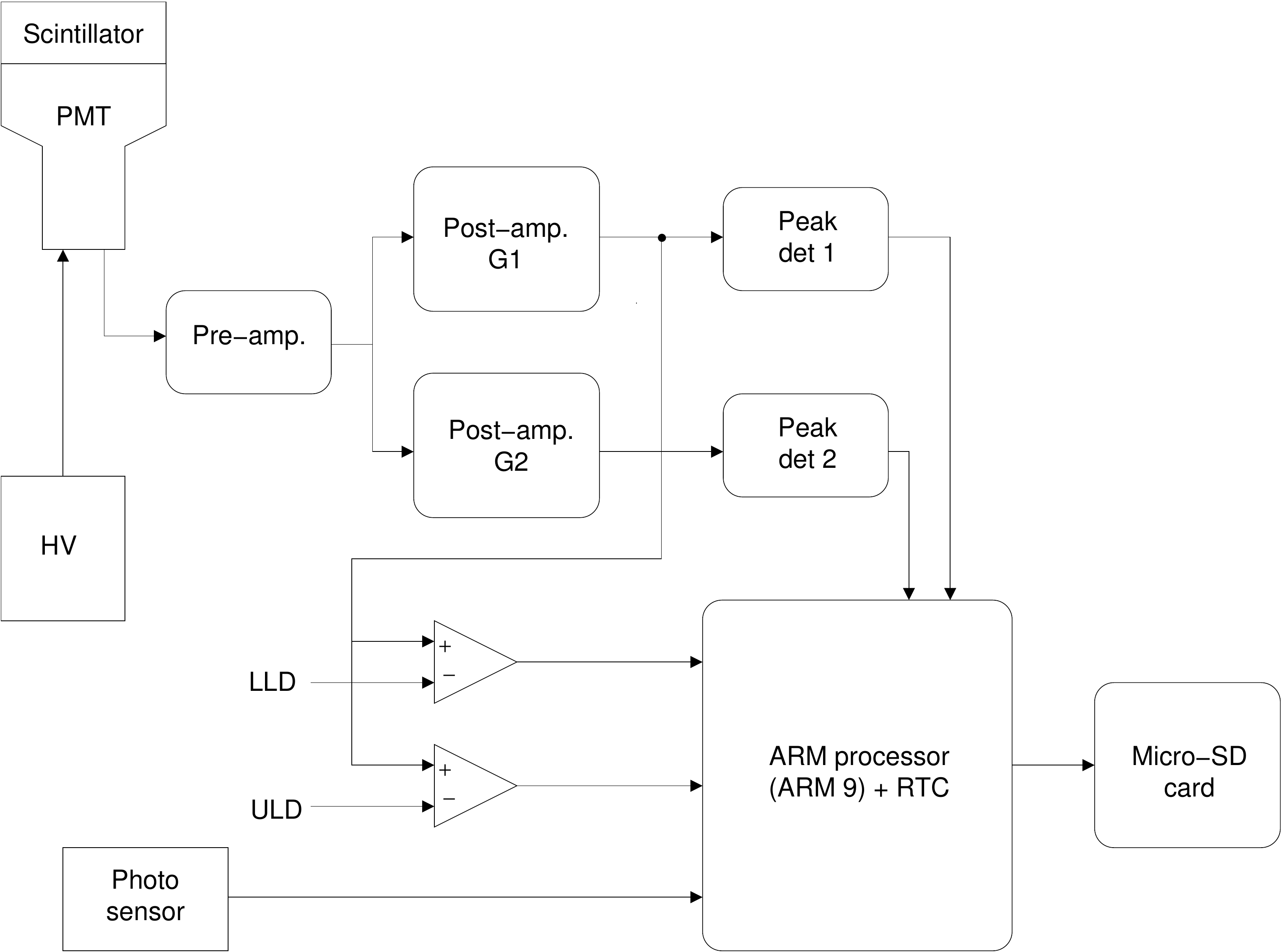}
  \caption{Schematic block diagram of the scintillator detector assembly.}
   \label{fig:scint}
\end{figure}

Fig. \ref{fig:scint} shows the block diagram of the whole scheme of the
signal readout system and the scintillator detector assembly.

Post-amplifier amplifies the pre-amplifier output pulses without affecting the
pulse shape (i.e., decay time). It also provides low impedance to the following
processing and analyzing circuit. Though we are generally interested in the
energy range of $\sim 15-150$ keV, still sometimes we are also interested in
higher energy phenomena (such as terrestrial gamma ray flash or TGF) and for this reason, in our circuit we have
the provision to read up to $\sim 2000$ keV. This large range of energy band
cannot be handled by a single amplifier without sacrificing the energy resolution. 
So, as shown in Fig. \ref{fig:scint} we
use two post-amplifiers: one of high gain (G1) for the energy range $\sim 15-150$
keV and another (G2) for $\sim 100-2000$ keV, a low gain amplifier. 

We generally operate the detector in the event mode, i.e., we record each event
with its pulse height and arrival time information, with a $1$ ms resolution in
the time stamp. Since we do not transmit scientific data, we have no limitation 
on the data acquisition amount. We store the data on board. This is a clear advantage over data storage 
process in satellite observations as the latter depends on the downlinking rate. 

The primary energy range of our detector, according to our scientific goal covers the
range from $\sim 15-150$ keV where we expect to observe the X-ray emissions from 
point like sources. To reduce the background counts we need to
use a collimator. The FoV of the collimator depends on our observation
purpose and also on the constraint on the payload weight. 
Since we do not have any provision of pointing the payload to a fixed object, 
the payload will rotate at different rates at a different height and
not necessarily always in the same sense. Thus in order to detect a specific target
for a considerable amount of time, the FoV should not be too narrow.
Considering these constraints we generally use $0.5$ mm lead sheets in different
configurations to build the collimator. Use of $0.5$ mm lead collimate effectively
$\sim 95\%$ up to $100$ keV. The use of lead as the collimator has a problem at
around $75$ keV due to its K$_{\alpha}$ emission line, this can be,
up to some extent, corrected during the analysis of the data by subtracting the
line. In some of the missions we desire more precise spectral data. For these we designed
the collimator with $1.0$ mm tin sheet along with $0.25$ mm copper (to absorb the
K$_{\alpha}$ line emitted from tin). 

The energy range allows us to study low energy radiations from the sun and compact objects,
as well as the secondary cosmic rays. Above $\sim 100$ keV our detectors become collimator-less
due to the transparency. So events such as gamma ray bursts or ubiquitous 
electron-positron annihilation lines could be seen without any problem. The lead emission line or the 
annihilation line can be used as calibrators, though we often use other calibrators on board as well.

\subsubsection{Geiger-M$\ddot{u}$ller counter}
\label{sssec:gmc}
\begin{figure}[h]
  \centering
  \includegraphics[width=0.6\textwidth]{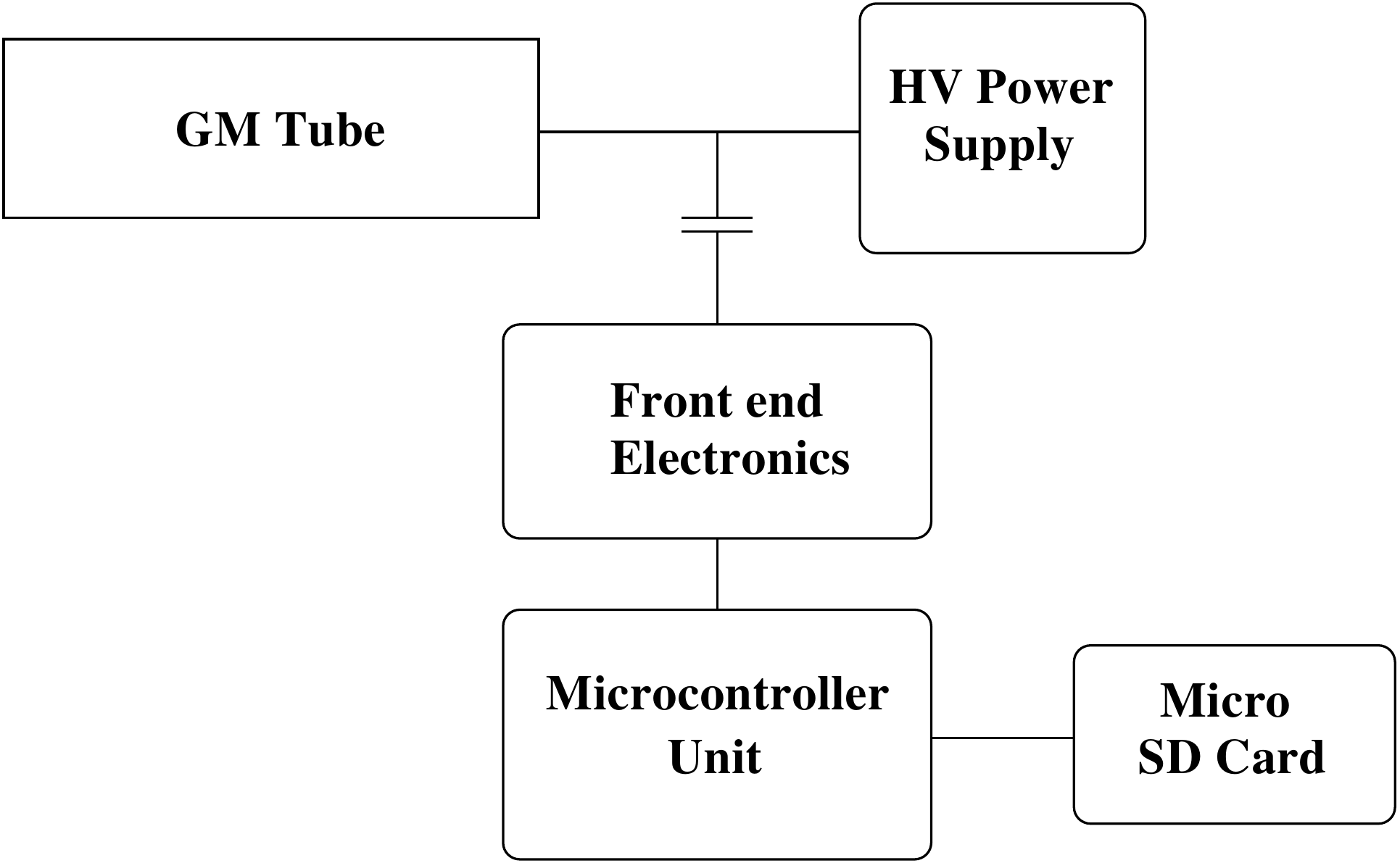}
  \caption{Schematic block diagram of the GM counter assembly.}
   \label{fig:gmc}
\end{figure}

Different kinds of GM counters were used mainly for the purpose of measuring the 
secondary CR in the atmosphere. One kind of these is a $4.9$ cm long and 
$1.5$ cm diameter Ne+Halogen filled cylindrical GM counter with stainless steel body and
a Mica window at one end. The other kind of GM counter is a $10.1$ cm long and 
$1.1$ cm diameter stainless steel tube filled with Ne+Halogen with no window. 
Sometimes we used several of these second kind of GM counter together to increase the 
effective detection area. These GM counters were used to count the secondary radiation
in the atmosphere using suitable readout electronics. A schematic block diagram
of the detector set up using GM counters is shown in Fig. \ref{fig:gmc}. These counters 
inside a 1cm lead tube were used also to measure atmospheric muons (e.g., Chakrabarti et al. 2013).

\subsection{Ancillary equipments}
\label{ssec:anci}

\subsubsection{Video unit}
\label{sssec:vdu}
Along with the purpose of capturing the extraterrestrial objects like Sun, moon, 
bright stars and planets, meteorites etc., the video camera unit
can capture an excellent areal view of the Earth from various altitudes. 
The unit acts as a visual monitor of the total mission, provides crucial information
about how the size of the balloons change with height, provides exact moments of
launching of payloads, balloon bursts and landing of the payloads apart from possible 
reason for entanglements between the balloon and the payload, if any, during the descend. 

\subsubsection{Sun Sensor}
\label{sssec:ss}
Especially, for the solar X-ray observation we use a Sun sensor device to cross-check if
the sun was indeed in the field of view when the data in MMU is sorted out as coming from the sun. 
For this purpose, we use a photo-sensor to capture the optical photons from the Sun. 
The FoV of the Sun sensor is kept to be the same
as the FoV of the detector in the MMU by using suitable collimator and aligning
the axes of the two instruments. 

To make the sensors light weight, the collimators are made from a black cardboard roll
held tight with duct tapes to provide mechanical strength. The photon 
intensity from the Sun at higher altitudes during the mission flight gets higher
when scattering of the lights becomes small due to rarefied atmospheric medium.
This saturates the photosensor. To reduce the photon intensity, filters are used in front of the
sensor. Filters made of two $2.5$ mm thick black acrylic sheet are enough for this purpose.

\subsubsection{GPS unit}
\label{sssec:gps}
GPS is one of the most important modules used to obtain instantaneous
payload location (attitude, longitude and altitude) and the wind
parameters at different levels through the atmosphere. The payload location
information is very important for the later use during the data analysis
as well as to track the payload for its recovery.

The GPS module used in the payload has a form factor of 
$19.0\times27.0$ mm and is of light weight. This module is connected with an 
external antenna. The receiver uses a patch antenna and provides serial data output.
The power consumption by the module is also negligible.

\subsubsection{Payload attitude measurement unit}
\label{sssec:dof}
Due to the limitation on the payload weight it is not possible to include a
conventional pointing device in the payload for continuous observation of any specific target.
Instead, we use a device to measure the payload attitude with time, so
that later during the analysis of the data we can extract the instantaneous
direction of the detector. We use Inertia Measurement Unit (IMU) sensors for this purpose.
The data from these sensors are processed by a processor board and stored in the
memory for future analysis.

The main components of the AMU are are followings:
\begin{itemize}
  \item One three axis Accelerometer
  \item One three axis Magnetometer
  \item One three axis Gyroscope
\end{itemize}

The 3-axis accelerometer provides the linear acceleration along its three axes.
The acceleration due to gravity (g) can be used as the reference to calculate
the tilt angle of the device with respect to earth. This is employed in
calculating the two Euler’s angles: the roll and pitch.

The 3-axis Magnetometer provides components of the Earth’s magnetic field
magnitude along its axes. This parameter is essential in computing the heading
angle for the payload.

The 3-axis gyroscope provides the angular rate around its each axis. The
accelerometer and the gyroscope together are used for sensor infusion and
calculation of exact roll and pitch angles. The pitch and the roll angle data is
then used for tilt-compensating the Magnetometer for accurate heading angle
calculation. However, bias, hard-iron and soft-iron calibrations are performed
prior to tilt-compensation during the ground test and calibration. During the
flight operation we only record the data from the sensors for all the axes
and store them with proper time tag. During data analysis we extract
the pitch, roll and yaw angle information of the payload using these data for a
particular time and consecutive altitude and azimuth (or, RA/DEC) of the
detector viewing direction.

The IMU sensor stick is connected to the Raspberry Pi processor board which is
the main computing unit here. This board gets the time information from an
external RTC module. Additionally, this board acquires other house keeping data
like pressure, temperature and location data from the pressure/temperature
sensors and GPS module. All these information are processed in sub-second time
interval and stored in an external SD card. 

\subsubsection{Pressure and temperature measurement unit}
\label{sssec:ptu}
We use two temperature sensors and one pressure sensor in our payload.
The pressure sensor and one temperature sensor are installed outside the
sealed payload box to measure atmospheric temperature and pressure and
one temperature sensor is used inside the box to monitor the temperature
of the motherboard inside. The pressure sensor is a piezoresistive material which when
subjected to some mechanical force provides a response in terms of
voltage change. Here, the force is exerted by air on the piezo
diaphragm of the sensor. The sensor produces its response which is
an analog quantity and fed to an analog to digital converter to convert
it into a form which the on board computer understands. The digitized
data is then stored to a solid state storage media on board. The
temperature sensor used is an integrated circuit. It provides a
change in current flow through it when subjected to temperature
change. A series resistor is employed with the transducer to
convert the current into a voltage which is fed to an ADC and
then saved in the storage media by the on board computing unit. The
calibration of the sensors is done in our lab with our setup.

\begin{figure}[h]
  \centering
  \includegraphics[width=0.6\textwidth]{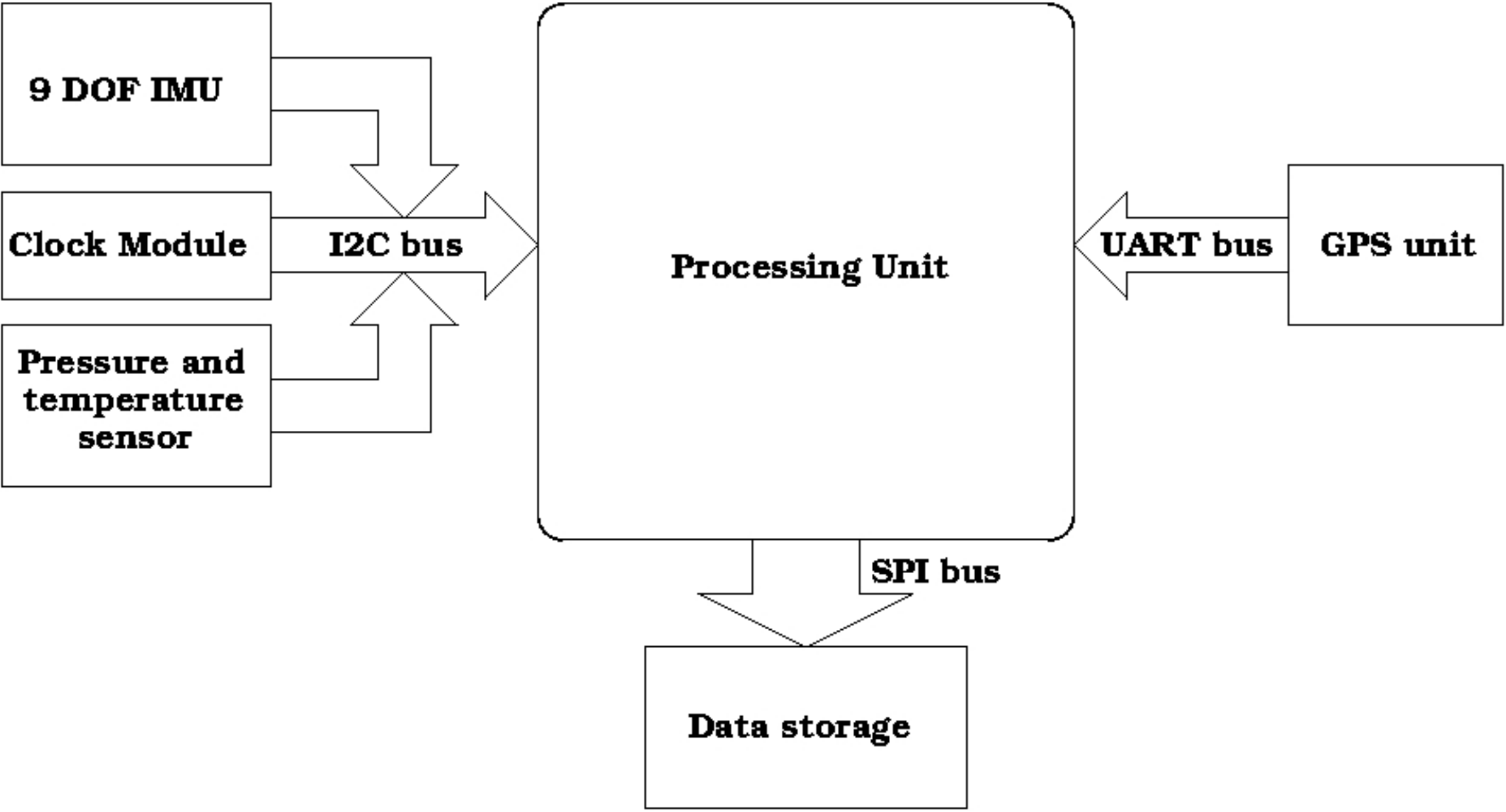}
  \caption{Schematic diagram of all the ancillary units used on board a typical payload.}
   \label{fig:allunit}
\end{figure}

Figure \ref{fig:allunit} gives 
a schematic diagram of how we combine data from various units and store on board.

\subsubsection{Positional Information Transmission Unit}
\label{sssec:pitu}
The recovery of the payload largely depends on the anticipation of its
landing location and on-the-go tracking features used in the mission. For this
purpose we designed a light weight and rugged device which can give the real
time positional information of a payload on board the carrier to the mobile
ground base. This live data of the payload location helps us to track the
current position of the payload and also to optimize the prediction of the
landing location.

The Positional Information Transmission Unit (PITU) on board in the payload
along with its receiving part in the mobile ground base comprise the payload
tracking system.

The transmitter unit in the payload consists mainly of three parts: GPS information
receiving unit, computing unit and the Transmitter. The GPS information 
is read in the system from the GPS module. A micro-controller is
used to implement the functionalities like parsing, preparation of the data
packet in proper format, synthesizing the signal data, transmitting the
data to the antenna via an amplifier. The whole part is implemented in firmware
of the micro-controller. We use standard short range two way devices which
are generally used for voice communication at the RF transmitter. The
micro-controller modulates the data with the two generated tones of $2200$ Hz and
$1200$ Hz. These modulated signals are transmitted from the payload to the ground. At
the ground station, the receiver tuned to the same channel, receives and feeds the output 
to a standard sound card of PC/laptop to digitize the analog signal. For
meaningful decoding of the received data, many publicly available packet radio decoder
software are available. This software unit decodes the signal packets. 
There are also provisions where one can 
directly use the data over a calibrated map (such as the Google map) 
to get the instantaneous position of the payload in the sky. 
There is also a provision to log the data. The data
obtained as the balloon goes up is used for predicting an estimated landing
location with an estimated landing path of the payload after balloon bursts. 
Of course, in case we miss some data in transmission due to signal weakness,
we can get the full data set from the GPS unit on board after payload recovery.

The requirement of the payload recovery has led us to implement another module in
the payload: Position Alert System Unit (PASU) to send the final location of
the payload upon landing. For this system we use a GSM/GPRS/GPS Tracker device
which are commercially available in the market (used for the purpose of vehicle
tracking, for example). This device works using existing GSM/GPRS network and GPS
satellites and is able to locate or monitor the remote target by SMS or
internet. It uses GPS, GSM dual position information and automatically stores the
GPS location information when the GSM signal is bad. This is a very compact and
light-weight module (dimension $64 \times 46 \times 17$ mm$^3$ and weights
only $50$ g), uses GSM/GPRS network for communication (using $850/900/1800/1900$ MHz
band). The position accuracy is up to $5$ m, so upon receiving the landing  coordinates
through SMS we can reach the exact position to recover the payload. The power
required by this device is very nominal and is supplied from the main power
supply of the payload. 
PASU sends out an SMS with the location information of the payload to the caller
when it is in contact with a cellphone network.

%% ------------------------------------------------------------------------ %%
\section{Test, calibration and responses}
\label{sec:calib}

\subsection{Tests in Atmospheric simulator}
\label{ssec:test}
During the flight, the detector undergoes a reasonable amount of pressure
and temperature change and we are restricted to use no pressure container or
better temperature shielding due to the weight limit on the payload. In order to 
test that the sensor would perform during the flight as per specifications, we need to 
test the detector and other payload components under such simulated environment. This 
enables us to calibrate the performance with varying atmospheric conditions.

For the testing of the payload components under temperature variation we use a
temperature simulation chamber where in a encapsulated vessel we gradually bring
the temperature down (to about $-10^{\circ}$C) from room temperature.
The pressure variation may affect the detector for any mechanical deformation
which may lead to the gain change of the detector. Another crucial effect is on
the electronics part especially on the high voltage biasing of the detector due
to the change of dielectric constant of the medium. Though we use a potting
material surrounding the high voltage bias part to prevent this effect, we
still need to test the system under low pressure condition. For this purpose, we
have a test bench containing an air-tight glass container fitted with suction
pump. This system can bring the pressure down to about $6-7$ mbar which
corresponds to about $35$ km height in the atmosphere. At this pressure value the
detector and other payload components show no deviation in performance.

\subsection{Detector calibration}
\label{ssec:detcal}
We have the detector data with a time resolution of $1$ ms and the whole energy
range is recorded in $1024$ channels. For calibration of the detector's energy 
response we use Eu$^{152}$, Ba$^{133}$ and Am$^{241}$ radiation 
sources which have $5$ lines inside our
energy range ($39.5$ keV and $121.9$ keV for Eu, $30.9$ keV and $81.0$ keV for Ba
and $59.5$ keV for Am). Fitting of these lines gives us a linear channel to 
energy relation. We also
calculate the energy dependance of the detector resolution which is useful for
spectral analysis of the data obtained from the detector. The payload on board
generally requires no calibrator since we have two natural ones: 
the emission line (at $\sim 75$ keV) from the lead collimator as well as the $511$ keV lines in high channels. 

\subsection{Detector and atmospheric response calculation}
\label{ssec:resp}
The balloons can take the payloads up to about $35-42$ km.
During the course, the detectors gather extraterrestrial X-ray data which have
to pass varying amount of residual atmosphere and hence suffer variable
absorption effects. So to trace back the original spectra of the X-ray data it is
necessary to calculate the absorbing nature of the varying residual
atmosphere. This can be done by means of the simulation with a proper
atmospheric model.

For this purpose, we consider the NRLMSISE-00 atmospheric model \citep{picone03}
to construct the geometry of the atmosphere. We simulate the propagation of the
photons and other charged particles through the atmospheric model by using
Geant4 simulation toolkit \citep{agos03}. 

Similarly Geant4 toolkit is also used for the simulation of the detector response.
We calculate the total detector response for the detectors with the shielding
and the collimator due to the incident photon energy from $10$ keV to $1$ MeV. This
response on the other hand will depend on the direction of the incident photons
due to the presence of the collimator, shielding material and geometry of the
detector. So we simulate the response of the detector for the photon in two
directions: parallel photon beams through the collimator aperture and isotropic
distribution of photons from the upper hemisphere of the detector coming into
the detector.

We also calculated the detector response for other major charged particles in
the atmospheric region of our interest which is also necessary to understand the data
from the detectors and as well as to extract the X-ray data from the background.
The detailed procedure will be discussed elsewhere.

%% ------------------------------------------------------------------------ %%
\section{Results and discussions}

We now present a few samples of our results to show the capabilities 
of our low-cost missions. In all our missions, we turn on 
the instruments one after another before the lift-off carefully recording 
each of the turn-on times in order to synchronize the data during analysis. They are 
turned off after descend. Thus, we always get the cosmic ray data 
in all our missions and large on board storage allows us to store this data for the entire trip. 

\subsection{Results from peripheral instruments}

\label{sec:res}
\begin{figure}[h]
  \centering
  \includegraphics[height=0.3\textwidth]{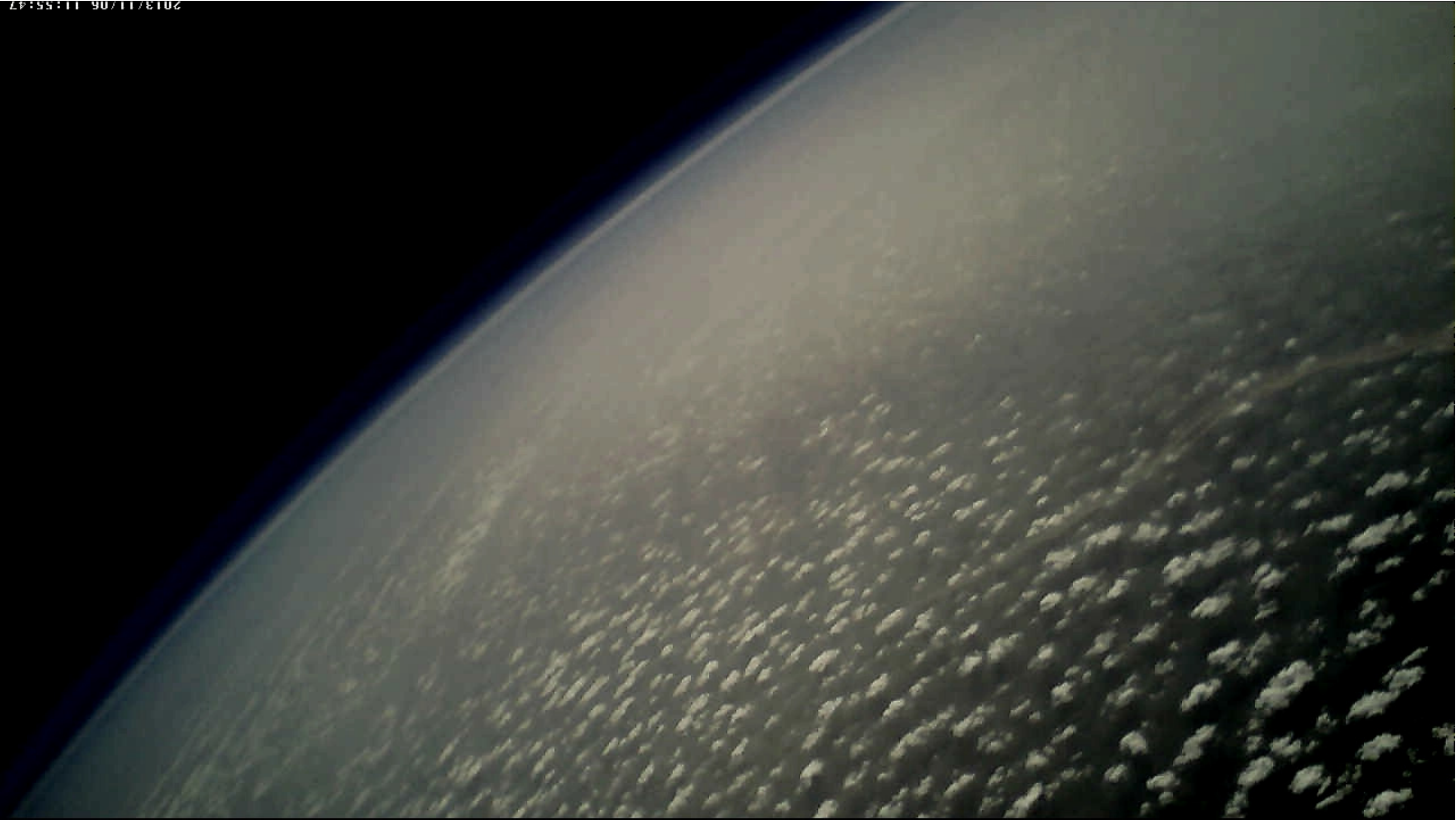}
\vspace{0.1cm}
  \includegraphics[height=0.3\textwidth]{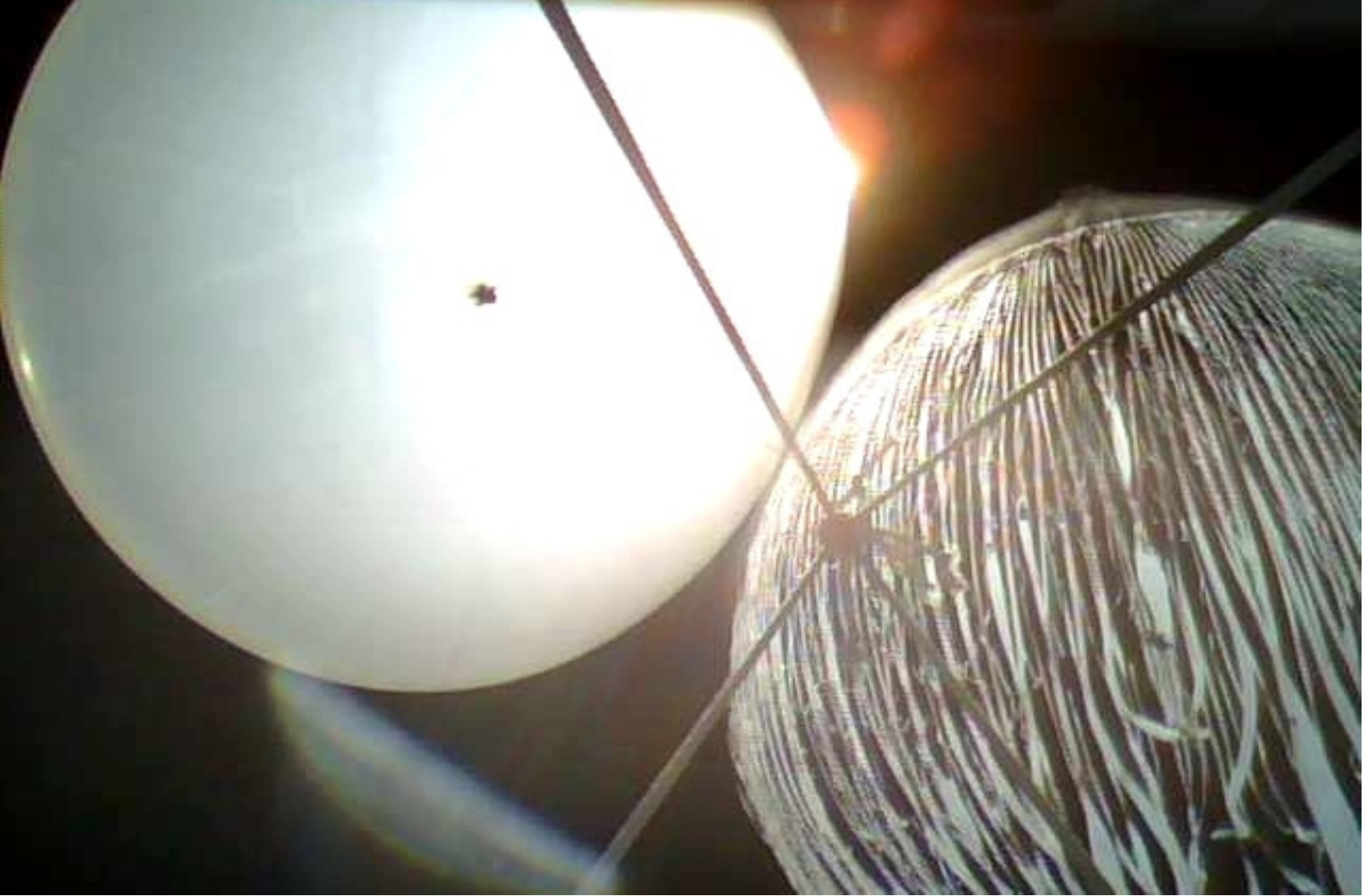}
  \caption{(left) A snapshot of the Earth from a payload at a height of $\sim 40$ km.
(right) Burst of one of the two rubber balloons while the other remains intact due to
lower lift given at the launching.}
   \label{fig:campic}
\end{figure}

First, we give two frames of a video taken from the payload in Fig. \ref{fig:campic}. 
On the left, we present a photo of the Earth from a height of $\sim 40$ km. 
On the right, we show the moment one balloon is shredded at its burst 
height. The other balloon remains intact. 

\begin{figure}[h]
  \centering
  \includegraphics[width=\textwidth]{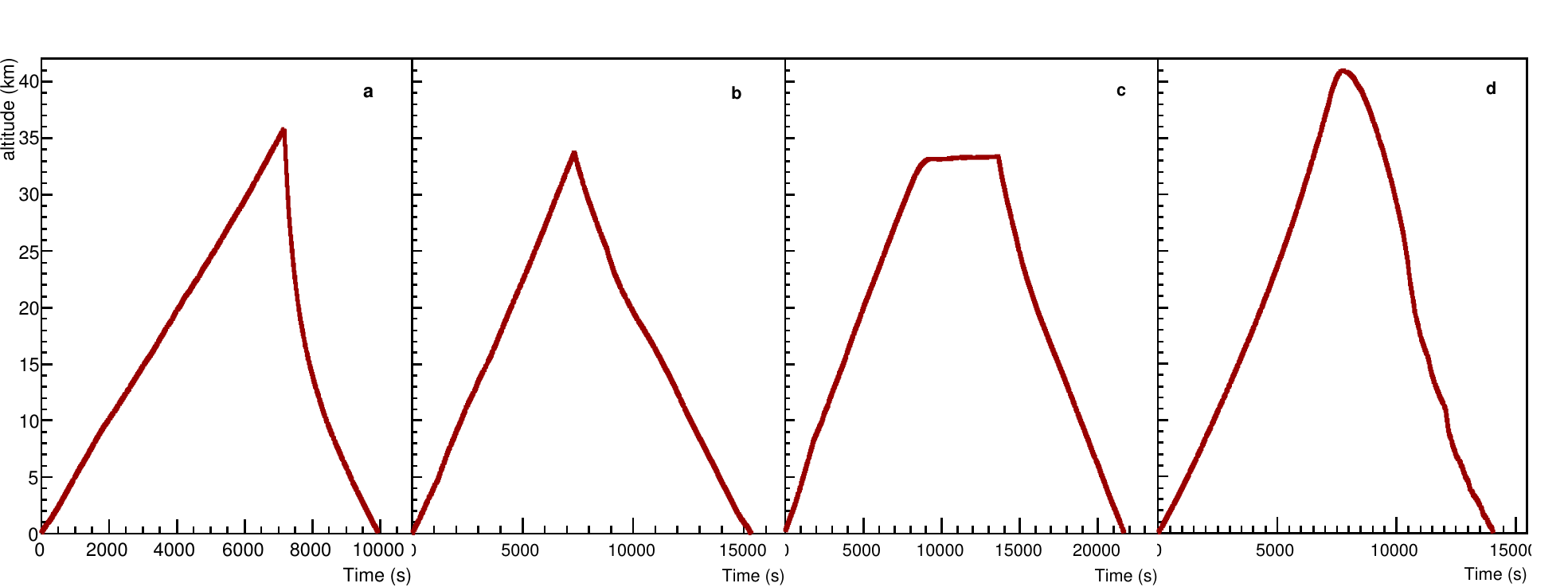}
  \caption{Payload altitude profile with time for various carrier modes (D75/90/52/92).
  (a) A single rubber balloon, (b) Two rubber balloons with heavier payload, 
(c)  A night flight with two rubber balloons, (d) A single polythene balloon.}
   \label{fig:tialt}
\end{figure}

\begin{figure}[h]
  \centering
  \includegraphics[width=0.6\textwidth]{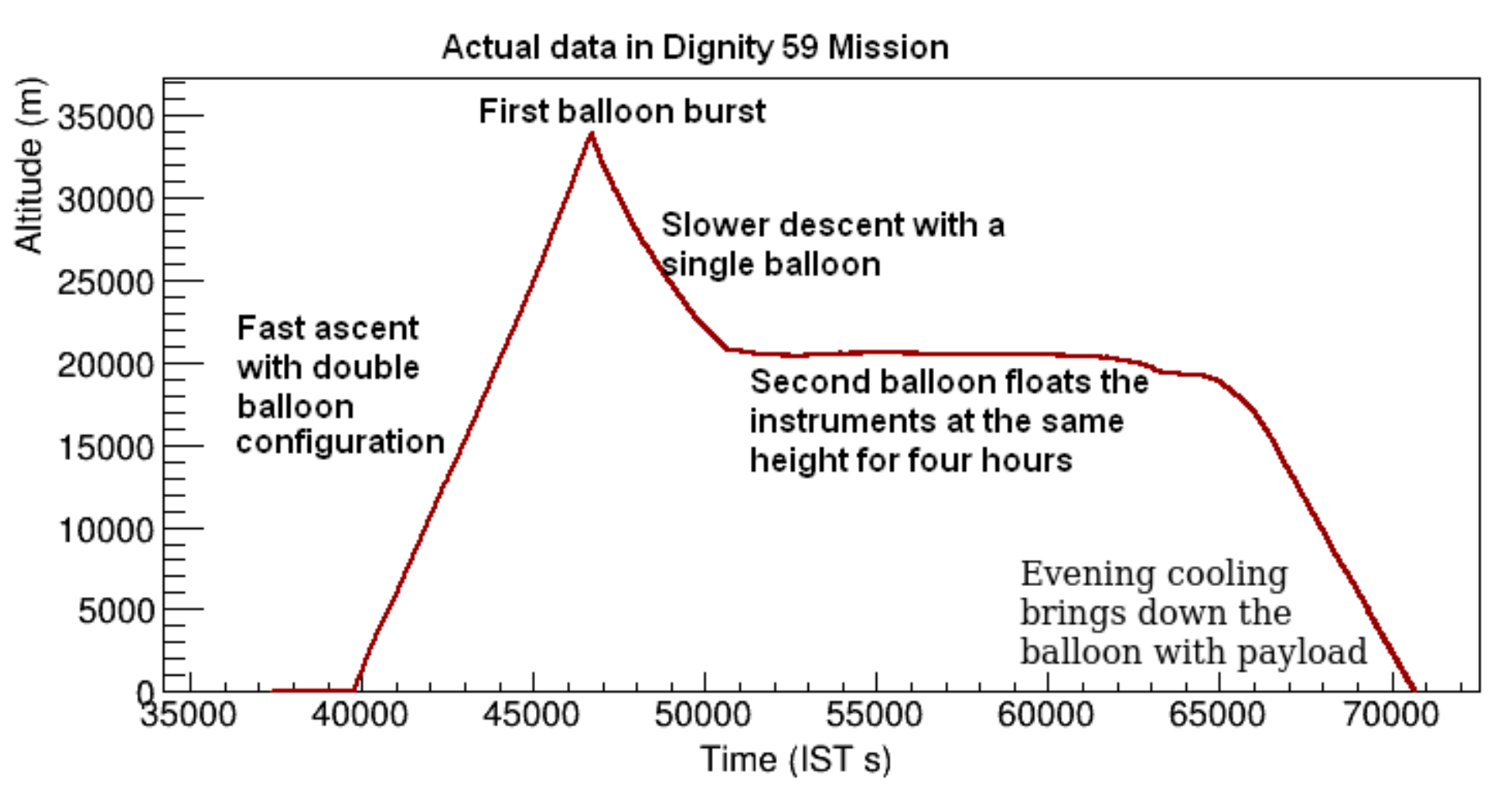}
  \caption{Time height profile during D59 mission. Near neutrality is achieved
at a height of about 20 km after a balloon burst. }
   \label{fig:d59th}
\end{figure}

As discussed before, the variation of altitude with time depends strongly on 
the lifts given at the time of launching and the combination of balloons. In our
experience of launching 101 Dignity Missions (D1 to D101), we have come 
across mainly five types of profiles four of which are given in 
Fig. \ref{fig:tialt} and the fifth with some description 
is given in Fig. \ref{fig:d59th}. In Fig. \ref{fig:tialt}(a), we show a typical (Dignity 75 or D75 mission) 
single time-altitude profile where the payload steadily reaches the burst height with almost constant 
vertical velocity. Right after the burst, the payload falls almost freely to a height of about $20$km 
and then as the parachute becomes steady, the landing is slower. In Fig. \ref{fig:tialt}(b),  taken 
from D90 mission, we see slower descend from the very beginning. Here, one balloon bursts, but the heavy 
payload continues to descend due to lower lift of the remaining balloon. In Fig. \ref{fig:tialt}(c), 
taken from D52 which was launched at a winter night, the payload height is saturated as the balloons 
shrink at lower temperature of the stratosphere. At the sun rise, they went up a bit till one bursts 
and the other brought the payload down. In Fig. \ref{fig:tialt} (d), taken from D92 where a polythene 
balloon was used, the ascending, descending as well as the behaviour above $\sim 30$ km are different 
from those of rubber balloons. We note that the peak is blunt as the balloon continues to rise even 
when it is torn apart at places. The descend is smoother as well.

In Fig. \ref{fig:d59th}, the time height profile during D59 mission is shown where the effect
of near neutrality due to burst of a single balloon is demonstrated. During the ascending phase,
the velocity was almost constant. After the burst of one balloon, the other brings the payload
down to an altitude of about 20km and floats at neutral height till evening sets in at Indian Standard
Time (IST) $\sim 64000$ s when the balloon shrinks and the lift is reduced. This finally brings 
the payload down to the ground slowly. In one of the Missions (D26) we achieved the mission
duration of about $12$ hours in this process.

\begin{figure}[h]
\centering
\includegraphics[width=0.6\textwidth]{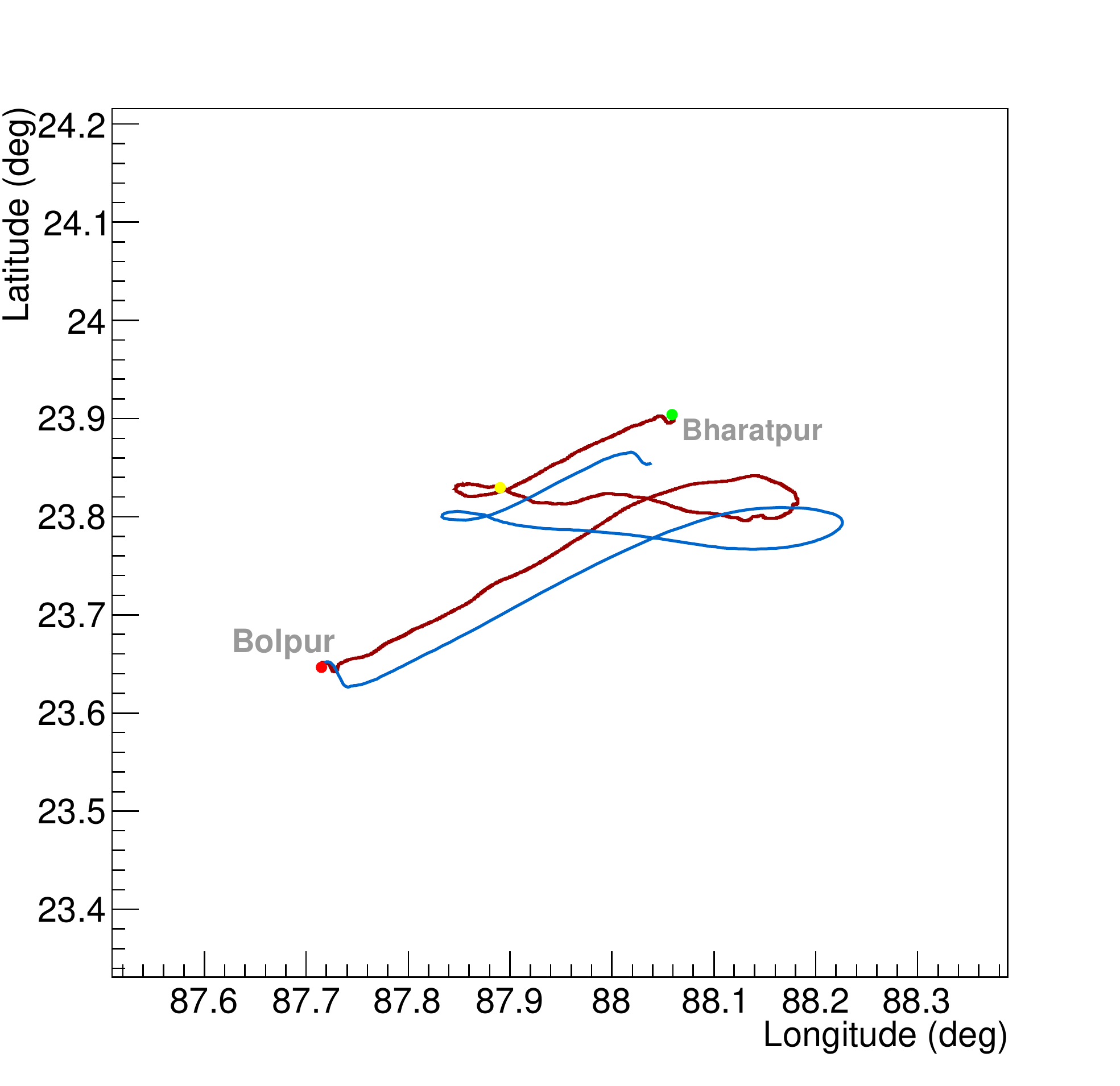}
\caption{Comparison of the predicted and real flight trajectories plotted on a google map 
of the region. The predicted data (blue) is taken from the online flight path predictor before
the mission with appropriate input parameters and the real data (red) 
from the Mission GPS data.} 
\label{fig:trajcomp}
\end{figure}

The most important aspect of our Missions is the recovery of the payload, as the 
scientific data is on board and the budget is limited. So, it is extremely 
important that we reach the landing site as the payload lands.
For this purpose, prediction of the payload trajectory with appropriate input parameters
and the selection of lifts, parachutes (to manipulate the descend rate) 
etc. before the mission is very crucial. We generally have a good predictability
rate. Very often we see the descend of the parachute (Fig. \ref{fig:para} in D17).
A comparison of the predicted and real trajectory of 
a mission for D91 is shown in Fig. \ref{fig:trajcomp}.

\begin{figure}[h]
  \centering
  \includegraphics[width=0.6\textwidth]{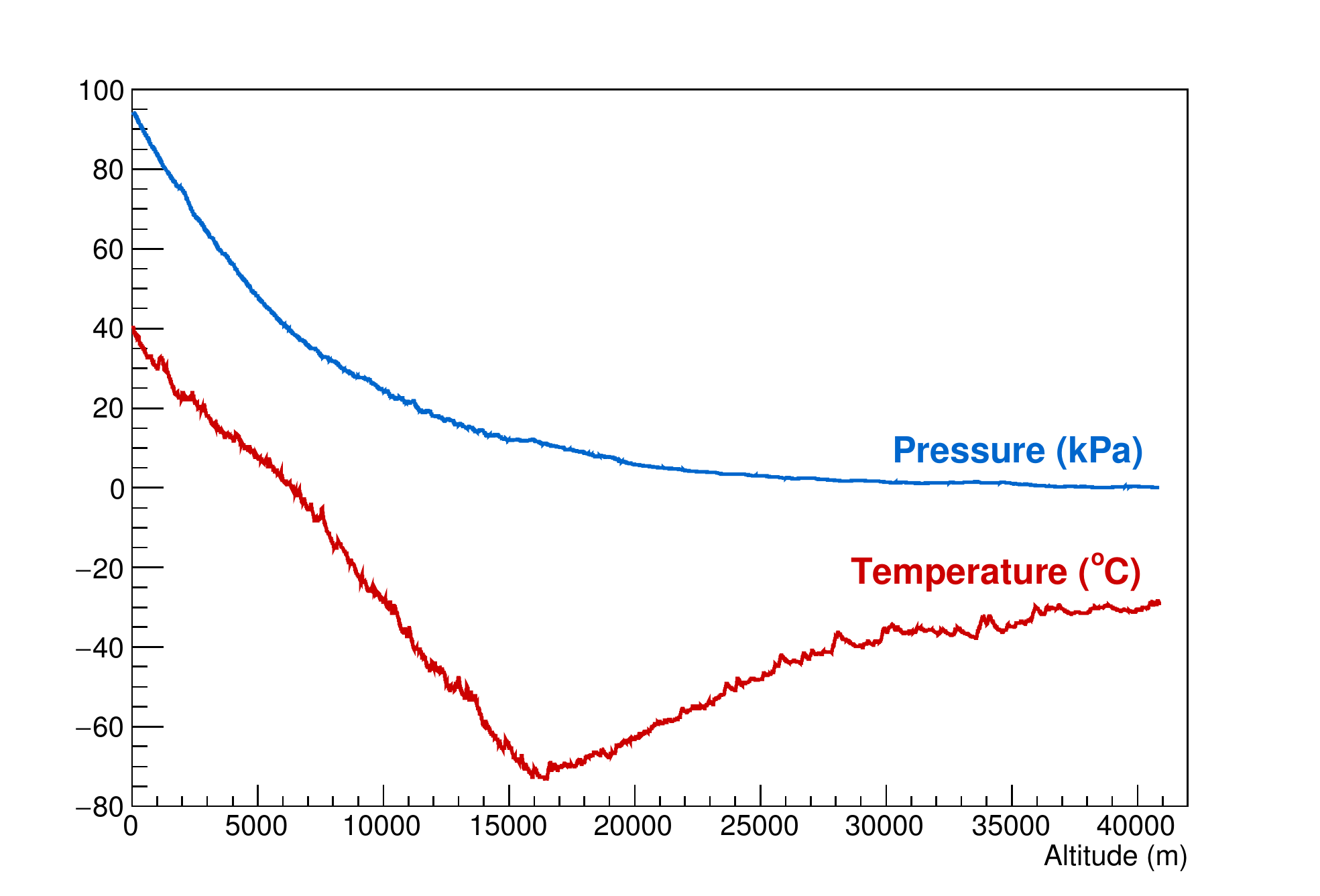}
  \caption{Atmospheric pressure and temperature profile with altitude in D97 Mission.
  The atmospheric temperature (in $^{\circ}$C)
  external to the payload is shown by red and the atmospheric pressure (in kPa) is in blue.}
   \label{fig:prte}
\end{figure}

In all our missions, we obtain the atmospheric parameters, such as
the pressure and temperature outside the payload box.
These parameters are important to check the working environment of
the detector. However, our data could be very useful for meteorological purposes
as we reach heights of $40$ km or more. The variation of pressure and temperatures with
altitude is shown in Fig. \ref{fig:prte}.

\begin{figure}[h]
  \centering
  \includegraphics[width=0.6\textwidth]{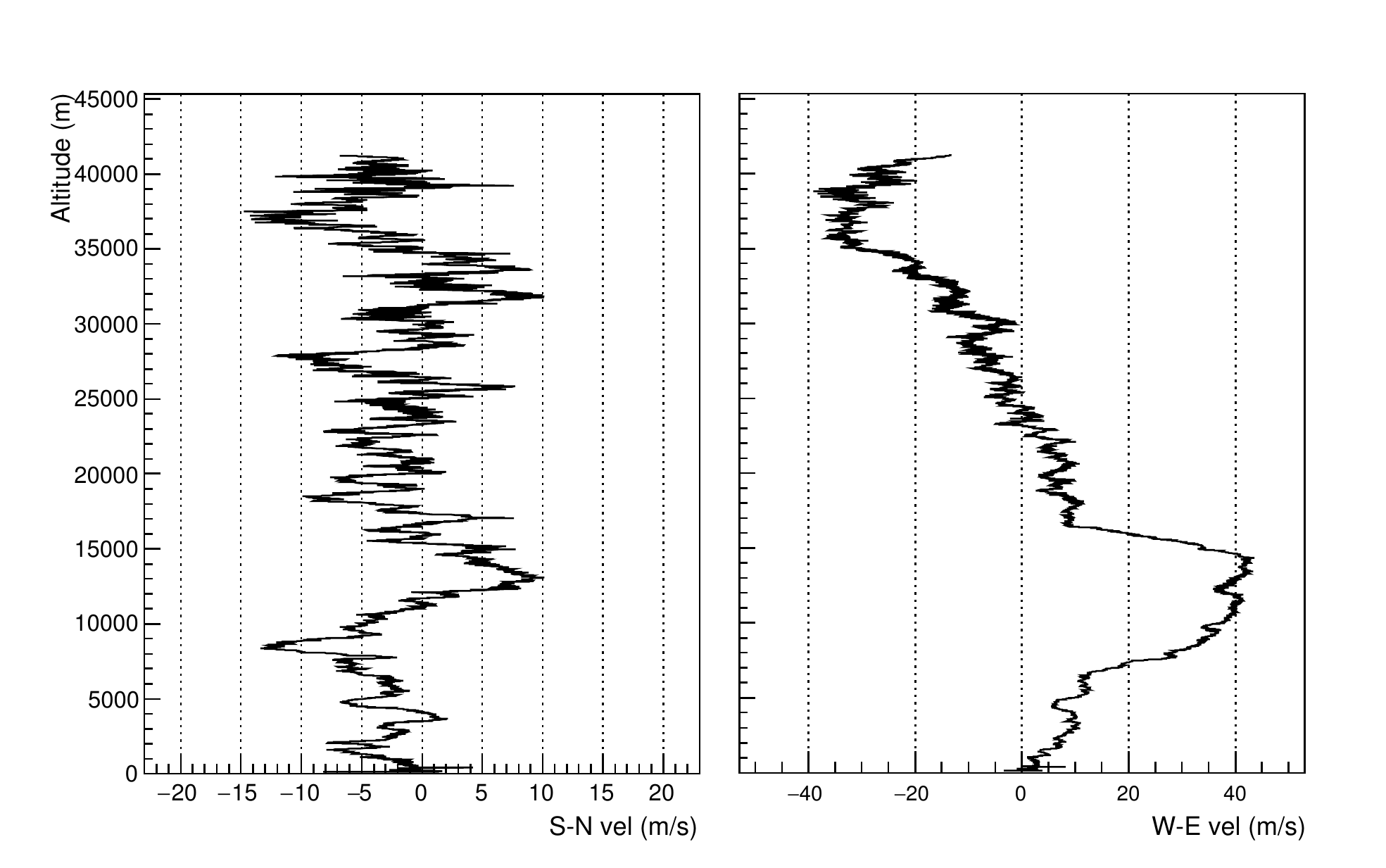}
  \caption{Wind velocity profile in the south-north and west-east directions
  at different altitudes up to more than $40$ km in D101.}
  \label{fig:wind}
\end{figure}

Measurement of wind velocity at different altitudes is important in 
the atmospheric study as well. Components of wind velocity is a byproduct from 
all our missions. We can retrieve the wind velocity profile up to a 
very high altitude which is not generally done in the regular atmospheric 
studies. As an example, we present the profile of the wind 
velocity in south-north and west-east directions at different
altitudes as shown in Fig. \ref{fig:wind}.

\begin{figure}[h]
  \centering
  \includegraphics[width=0.6\textwidth]{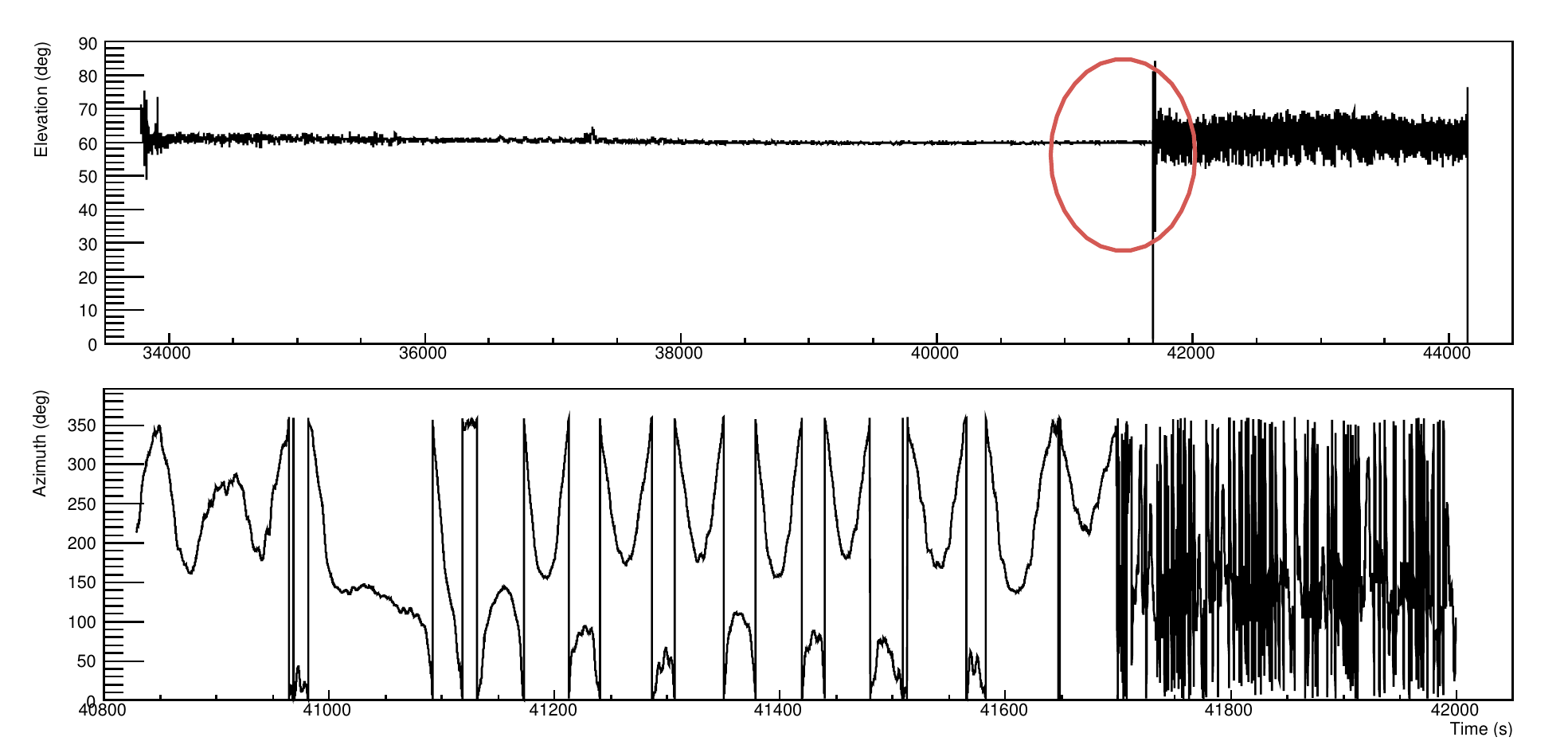}
  \includegraphics[width=0.3\textwidth]{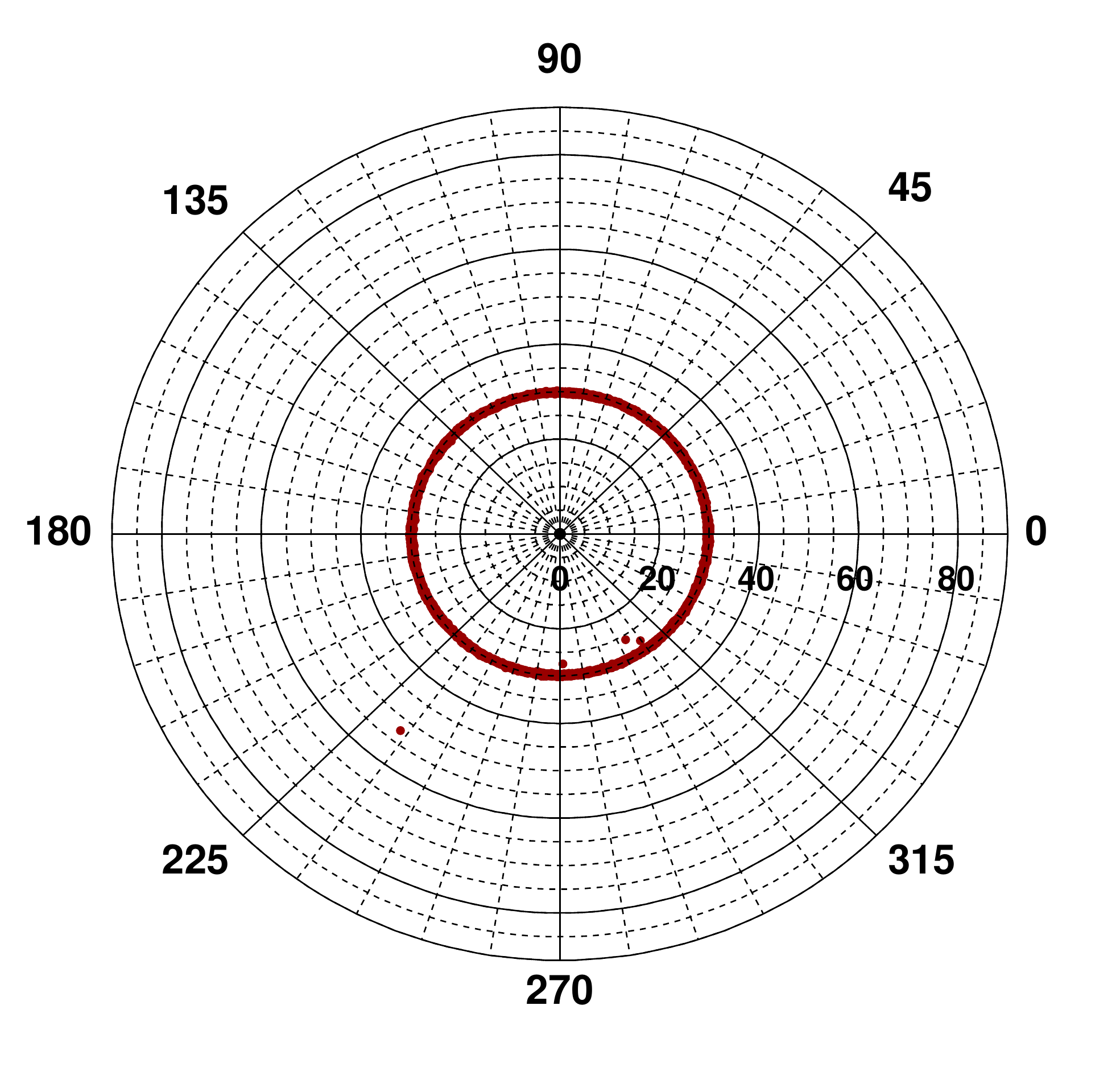}
  \caption{Payload attitude information obtained from the IMU in D98.
Two panels on the left show elevation (top panel) and azimuth angles (bottom panel) around
the burst time marked by an ellipse of the detector direction (axis of the MMU). On the
right we show the annular part of the sky covered by the detector direction vector (red points) 
above 30 km altitude till burst.
}
\label{fig:att}
\end{figure}

Depending on the target source, the payload is tilted at an angle with vertical axis. 
However, the azimuthal angle is not known as priori. So,
the information of the payload orientation is important to know the direction from which 
photons are entering into our instruments. This enables us to locate the position 
of X-ray sources in the sky. We calculate the detector axis direction 
using the data of the AMU installed in the payload box.
In Fig. \ref{fig:att} we show the direction of the detector axis 
in horizontal coordinate system.

\begin{figure}[h]
  \centering
  \includegraphics[width=0.6\textwidth]{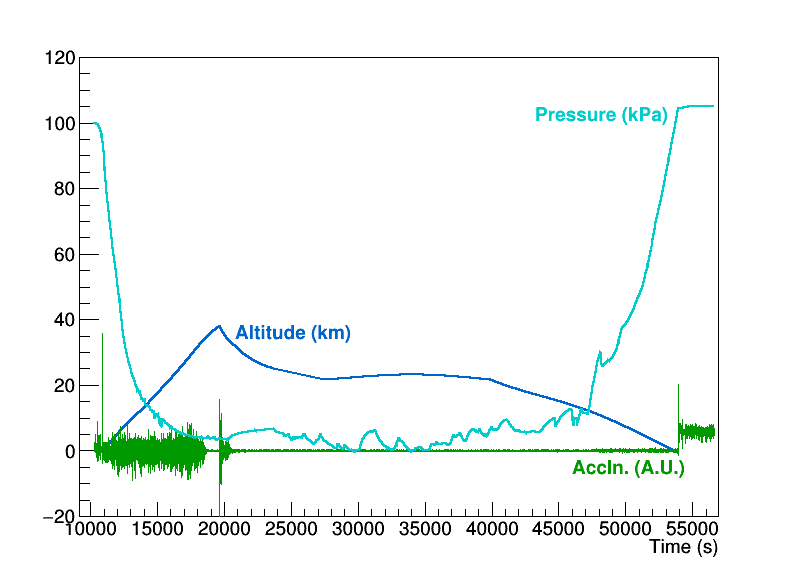}
  \caption{Pressure, Altitude and acceleration data of D26 mission showing a $12$ hour
  flight duration in a double balloon launch.}
   \label{fig:d26}
\end{figure}

Normally a single balloon flight lasts of about three hours. However, when a double
balloon is launched, depending on the distribution of the lifts and the mass of the 
payload, the duration could be very long, perhaps 12 hours or longer. One such 
mission parameters are shown in Fig. \ref{fig:d26} where we show the altitude variation in blue,
the pressure variation (kPa) in cyan and the IMU data (acceleration in x-direction) in green. Note that the payload
floated for almost five hours at about a constant height of 22 km. Note also the IMU parameter 
indicating the burst of a single balloon (spike at $\sim 20000$ s) and some turbulence thereafter till the neutrality 
height is reached. 

\subsection{Scientific data}

We now turn our attention to obtain samples of scientific data from our low cost
Dignity Missions. We present results of Cosmic Rays, Solar flares electron-positron 
lines just to indicate the data quality. Detailed data analysis  of specific missions
would be presented elsewhere. 

\subsubsection{Cosmic Rays}

\begin{figure}[h]
  \centering
  \includegraphics[width=0.8\textwidth]{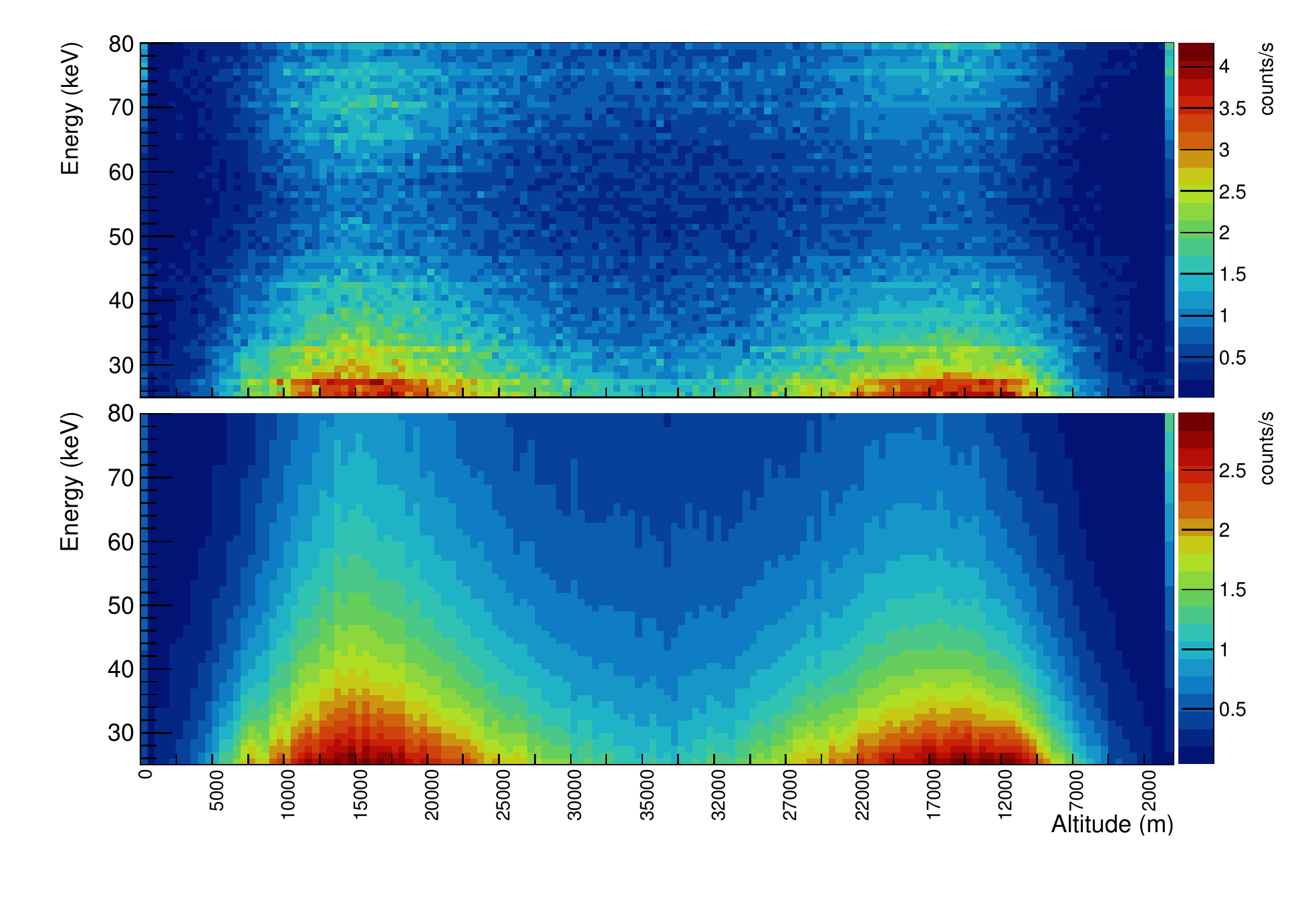}
  \caption{Dynamic spectrum of the secondary CR with altitude. The upper panel
  shows the detector data including emission effect from the lead collimator 
  shielding. Lower panel data shows the result after removing the emission effect.
  }
  \label{fig:crspec}
\end{figure}

The interaction of the primary CR with the atmospheric nuclei produces 
secondary cosmic-ray shower \citep{grie01,gais90}. 
Our detector is capable of detecting low energy X-ray 
part of these secondary products. Intensity of the secondary X-ray depends on 
the altitude as the atmospheric density and composition varies with altitude.
The interaction of the primary CR gradually intensifies at lower heights producing 
more secondary particles. But the loss effect due to absorption and decay processes
also become more at lower heights as the density gets higher. Due to the balance 
of these interactions we get a maximum secondary intensity around $15-20$ km 
depending on the geomagnetic latitude of the location \citep{bazi98, harr14, li07, shon77, yani16}.
In Fig. \ref{fig:crspec} we have shown the spectral change of the secondary CR dynamically 
with altitude.

\begin{figure}[h]
  \centering
  \includegraphics[width=0.6\textwidth]{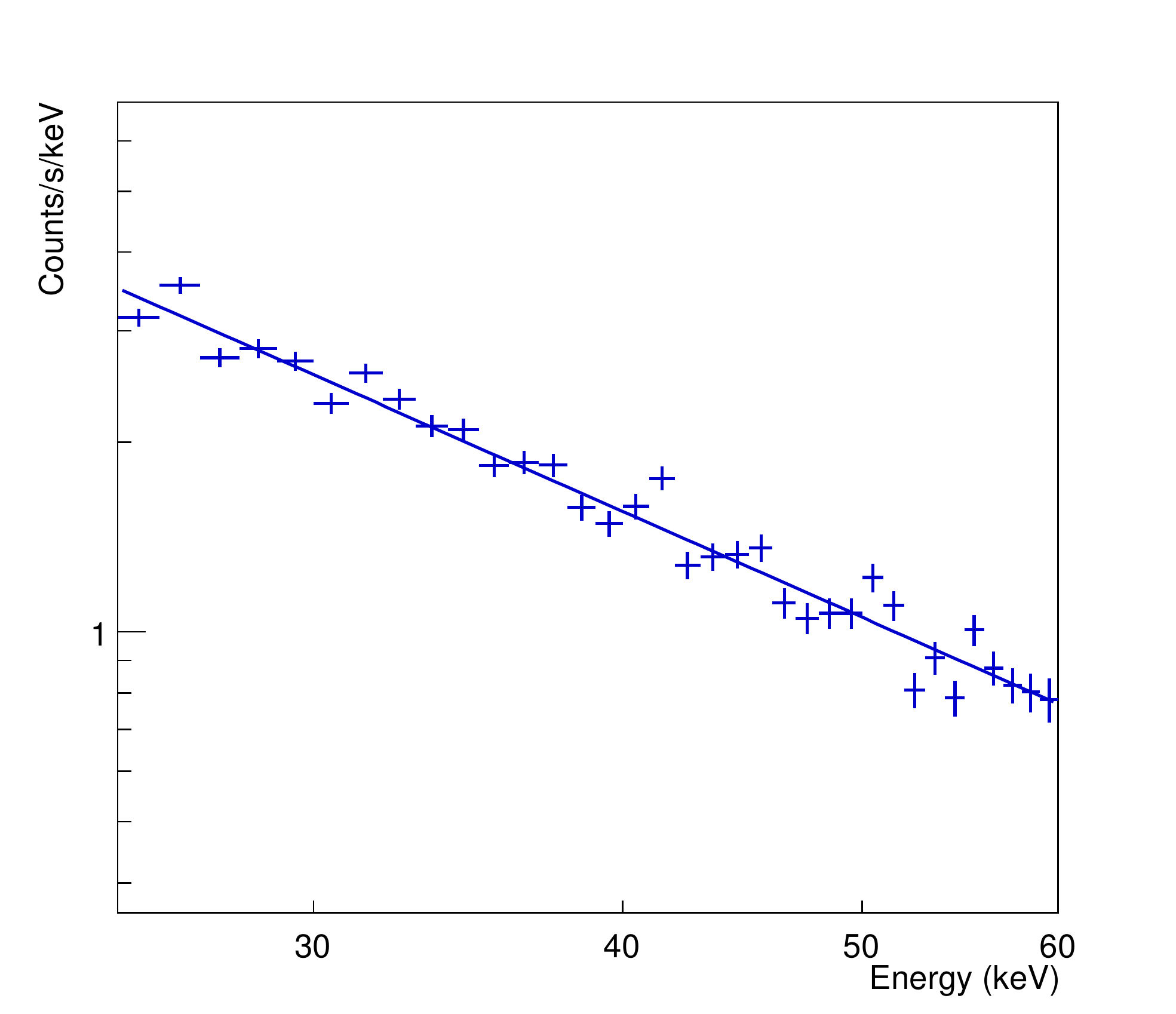}
  \caption{Spectrum of the secondary CR near the Pfotzer maximum. Solid line 
  is the best fit power-law curve for the spectrum.}
  \label{fig:sppfot}
\end{figure}

The secondary cosmic ray spectrum due to the interaction of primary cosmic rays
in the atmosphere near the Pfotzer maximum is shown in Fig. \ref{fig:sppfot}.
This spectrum shows a power law distribution of the form $931.7 \times E^{-1.73}$ 
shown by the fitted solid line along with the spectrum.

\subsubsection{Solar flare}

\begin{figure}[h]
  \centering
  \includegraphics[width=0.6\textwidth]{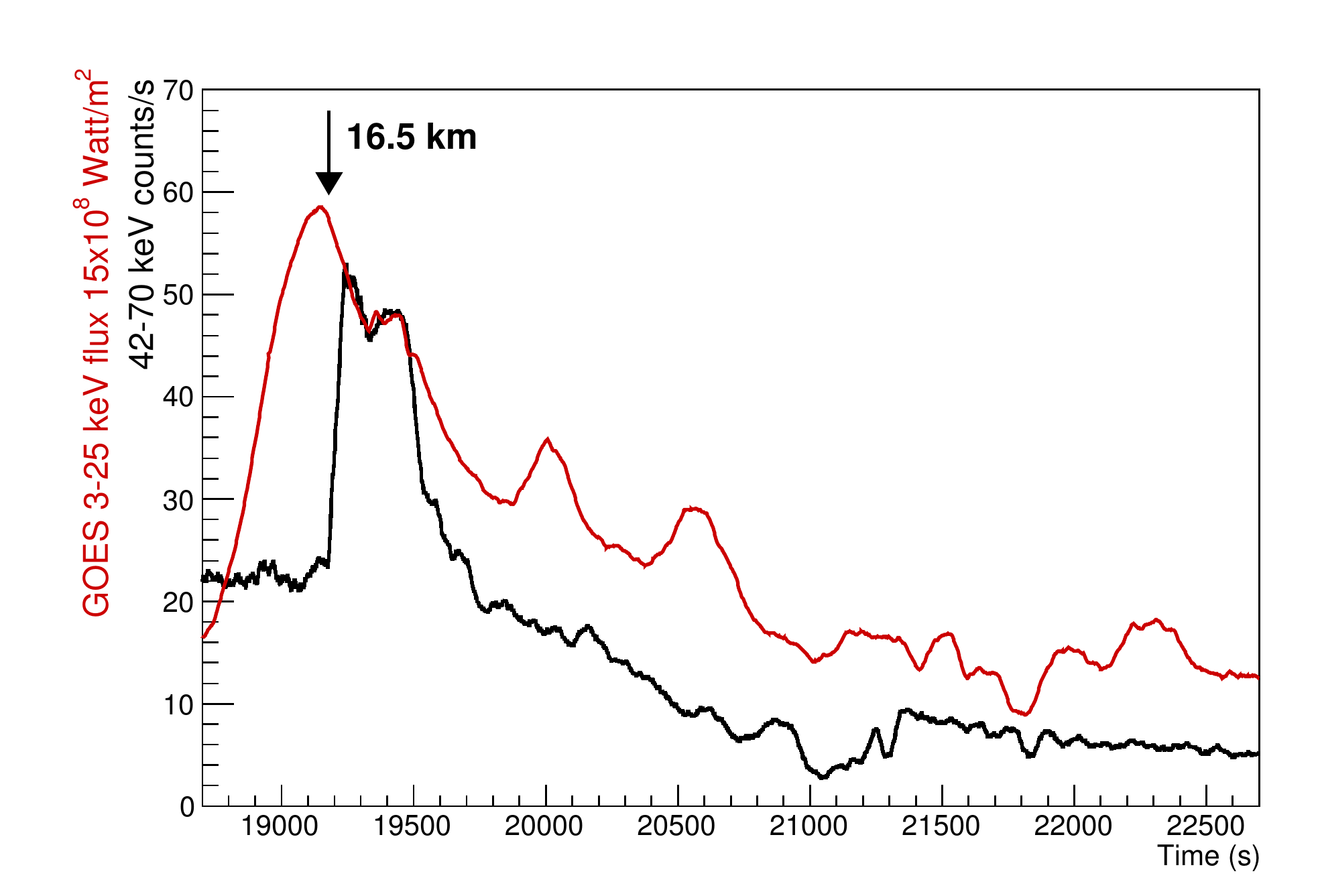}
\caption{Solar light curves in $42-70$ keV (black) showing the high energy detection 
of a solar flare during D40 mission on May 18, 2013 by a scintillator detector 
of $2" \times 2"$ NaI crystal. The $3-25$ keV solar flux data from GOES satellite 
also detected the same flare.} 
\label{fig:sollc}
\end{figure}

Since the Sun is the most X-ray bright extraterrestrial object in the
sky, we targeted it in several missions, especially during flare times. Our mission preparedness
allows us to prepare and launch as soon as the Sun becomes active. Only 
delay arises from the Airport Traffic Controller's permission to fly. Thus we have 
a good amount of data from the quiet and the active sun.
Fig. \ref{fig:sollc} shows the light curves in $42-70$ keV (black) 
during a solar flare. The Sun entered inside our FoV ($35^{\circ}$) of the $2"\times2"$
NaI(Tl) Bicron detector \citep{saint} after the payload ascended above the height of
the Pfotzer maximum. We plot the low energy ($3-25$ keV) light curve of NASA GOES satellite
\citep{swpc} also on the same plot 
(red). We lost the low energy data due to heavy absorption below Pfotzer maximum.
Our light curve generally follows the GOES light curve. However, high energy emission \citep{kruc08} 
is always of shorter time duration as we find here also. Lower intensity superposed GOES flaring events 
could not be detected at higher energy channels in our observation. This light 
curve is not corrected for changes in atmospheric absorption with height.

\subsubsection{$e^- - e^+$ annihilation lines}

\begin{figure}[h]
  \centering
  \includegraphics[height=0.3\textwidth]{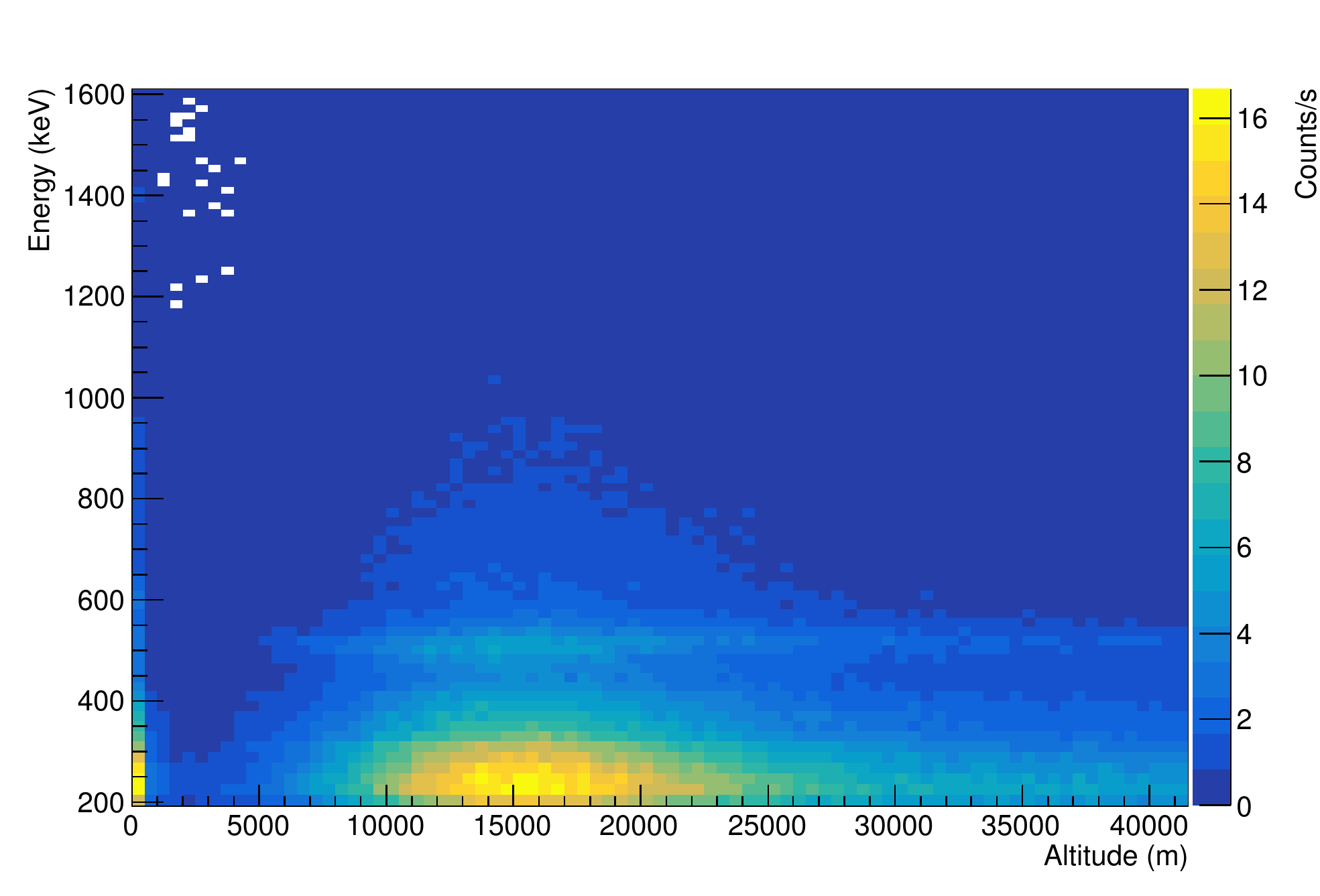}
  \includegraphics[height=0.3\textwidth]{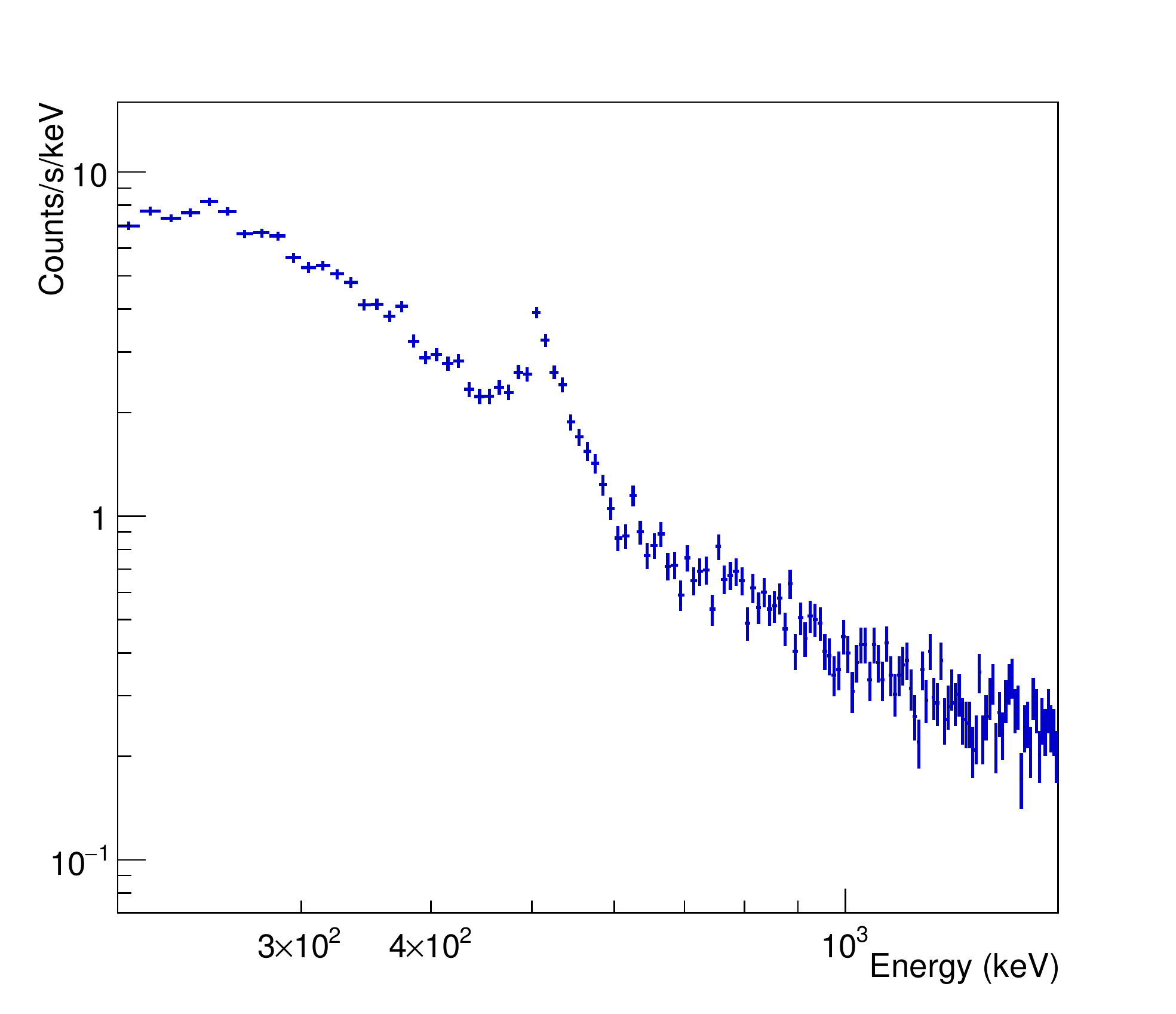}
  \caption{ (Left) Dynamical spectrum of secondary Cosmic Rays 
in high energy channels in D101 showing Pfotzer maximum at around 16.5km.
We also see the $e^-$ - $e^+$ annihilation line at 511keV. (right) 
Spectrum at the Pfotzer maximum showing clearly the annihilation line.  }
   \label{fig:spec511pfot}
\end{figure}

Primary and secondary cosmic rays impinging on the collimator or nearby metal casings will produce 
copious charge particles. Most importantly, electrons and positrons annihilating nearby are picked 
up by our detector and its strength is directly related to the secondary 
radiation intensity. In Fig. \ref{fig:spec511pfot}, left panel, we present the dynamic spectrum at very high energy (200 - 1600 keV)
as  a function of the altitude. The Pfotzer maximum can be seen at about $\sim 16.5$ km.  The $e^- - e^+$
annihilation line at 511 keV could be seen very clearly. The intensity of the line is proportional to the 
radiation content in the surroundings. At about $40$ km, the secondary radiation is very low and therefore 
this line intensity is also reduced. In the right panel, we plot the spectrum only at the Pfotzer maximum. The sharp 
line could be seen. 

\section{Conclusions}
\label{sec:conc}
In this paper, we discussed a new paradigm of near space exploration 
where one can carry out sophisticated space based observations using very low cost 
rubber and small polythene  balloons which are commonly used for meteorological purposes.  
The light weight payloads  contain both the main measurement unit with on board data
storage, as well as ancillary units such as GPS data transmitter, inertial measurement units, 
video cameras, sun sensors etc. Power is supplied from rechargeable Li-Po type batteries. We
concentrate on the X-ray sky, namely, the cosmic rays, the Sun and the compact objects.
Duration of each mission could be between 3 to 12 hours. Because of event mode data storage, 
we can save time stamps of each received photon. Inertia measurement unit enables us to tag each photon
with the directional information from which each photon is received. 
That way, even if we have no 
pointing device, we can separate photons from each source. 
We have shown that we have been able to obtain dynamical spectrum of  cosmic rays, solar flares, 
and even electron-positron annihilation lines have been detected quite prominently. We 
have presented wind and atmospheric parameters. In our double-balloon approach, we have been able to 
achieve a remarkable feat to float a payload at near constant height for hours by suitably 
manipulating the lifts of the balloons. We have presented qualitative pictures of how this
could come about.

Though there are many limitations of our low cost flights, we also have several advantages
which are not easily achieved by more expensive methods. For instance, we have 
Cosmic ray variation data year after year which appears to be anti-correlated
with solar activity \citep{sark17}. 
Because payloads are very light, we have taken many innovative 
approaches as each stage. The recurring cost is so low, effectively the cost of the 
balloons and the gas along with logistics, that younger generations can have access to 
serious space research. Our method is also applicable for testing small payloads
in preparation for future space missions. 
It is also useful to assess radiation hazards and aviation safety
\citep{miro03,icsp}.

Our main goal in the paper was not to present specific results, but to impress what could be
achieved, in principle. In separate papers, we will discuss in details of how X-ray spectra
of the Sun, or pulsar or black hole candidates could be obtained or the secondary Cosmic rays
could be monitored. We have procured an wealth of information about the pattern of atmospheric
parameters including pressure, temperature and winds before and after the Indian monsoon year
after year. These will be valuable for weather prediction. 

%% ------------------------------------------------------------------------ %%
\begin{acknowledgements}
The authors would like to thank Dr. S. Mondal, Mr. S. Chakraborty, Mr. S.
Midya, Mr. H. Roy, Mr. R. C. Das, Dr. S. Chakrabarti and Mr. U. Sardar for their valuable help
in various forms during the mission operations and data collection. We also
thank Ministry of Earth Sciences (Government of India) for partial financial
support.
\end{acknowledgements}

%% ------------------------------------------------------------------------ %%

%% ------------------------------------------------------------------------ %%
\end{document}